\documentclass[11pt,a4paper]{article}
\pdfoutput=1
\usepackage{jheppub}
\usepackage{graphics}
\usepackage{stmaryrd}
\usepackage{amsfonts}
\usepackage{mathrsfs}
\usepackage{color}

\title{Thermal Dynamic Phase Transition of Reissner-Nordstr\"{o}m Anti-de Sitter Black Holes on Free Energy Landscape}

\author[a,b]{Ran Li,}

\author[c]{Kun Zhang,}

\author[d]{Jin Wang\footnote{Corresponding author}}

\affiliation[a]{Department of Physics, Henan Normal University,
 Xinxiang 453007, China}

\affiliation[b]{Department of Chemistry, State University of New York at Stony Brook, Stony Brook, NY 11794-3400, USA}

\affiliation[c]{State Key Laboratory of Electroanalytical Chemistry, Changchun Institute of Applied Chemistry, Chinese Academy of Sciences, Changchun, China, 130022}

\affiliation[d]{Department of Physics and Astronomy, State University of New York at Stony Brook, Stony Brook, NY 11794-3400, USA}

\emailAdd{liran@htu.edu.cn}
\emailAdd{zhangkun@ciac.ac.cn}
\emailAdd{jin.wang.1@stonybrook.edu}

\abstract{We explore the thermodynamics and the underlying kinetics of the van der Waals type phase transition of Reissner-Nordstr\"{o}m anti-de Sitter (RNAdS) black holes based on the free energy landscape. We show that the thermodynamic stabilities of the three branches of the RNAdS black holes are determined by the underlying free energy landscape topography. We suggest that the large (small) RNAdS black hole can have the probability to switch to the small (large) black hole due to the thermal fluctuation. Such a state switching process under the thermal fluctuation is taken as a stochastic process and the associated kinetics can be described by the probabilistic Fokker-Planck equation. We obtained the time dependent solutions for the probabilistic evolution by numerically solving Fokker-Planck equation with the reflecting boundary conditions. We also investigated the first passage process which describes how fast a system undergoes a stochastic process for the first time. The distributions of the first passage time switching from small (large) to large (small) black hole and the corresponding mean first passage time as well as its fluctuations at different temperatures are studied in detail. We conclude that the mean first passage time and its fluctuations are related to the free energy landscape topography through barrier heights and temperatures.}

\begin{document}

\maketitle

\section{Introduction}

Hawking's original derivation of blackbody radiation from a collapsing black hole \cite{Hawking} revealed the thermal nature of black holes. This discovery established a close relationship between gravity, thermodynamics, and statistical physics. Since then, studying black hole from the thermodynamic and statistical physics viewpoint has attracted significant attention. Hut studied the phase transitions of charged black holes \cite{Hut}. Davies found that Kerr-Newman black holes can undergo a second order phase transition at the point where the heat capacities are divergent \cite{Davies}. By treating black hole as a state in the thermodynamic ensemble, Hawking and Page \cite{HawkingPage} studied the first order phase transition from thermal AdS space to the large AdS black hole at a certain critical temperature. More recently, the first order phase transition between the small and the large RNAdS black holes was studied by Chamblin et al \cite{Chamblin1,Chamblin2,Wu}. By treating the cosmological constant as thermodynamic pressure and the black hole mass as enthalpy \cite{Kastor:2009wy}, Dolan also discovered the remarkable analogy between van der Waals liquid-gas system and RNAdS black hole in extended phase space \cite{Dolan1,Dolan2}. Kubiznak and Mann \cite{Mann} showed the critical behavior of RNAdS black hole in fixed charge ensemble coincides exactly with that of van der Waals liquid-gas system.

The discovery of the anti-de Sitter/conformal field theory (AdS/CFT) correspondence \cite{Maldacena,GKP,Witten} greatly promotes the studying of phase transition of black holes in AdS space. It is well known that, by AdS/CFT correspondence, Hawking-Page transition can be properly explained as the confinement/deconfinement transition in quantum chromodynamics (QCD) \cite{Wittenphase}. In particular, the analogy between AdS black holes and van der Waals liquid has been studied extensively, including the small-large black hole transition in modified gravity and higher derivative gravity \cite{Cai,WeiprdGB,Zou,RKM,Mo,Hu,Fernando,Yazdikarimi:2019jux,Dehyadegari:2018pkb,Dayyani:2017fuz,Dehyadegari:2016nkd}, Maxwell equal area rule of van der Waals type phase transition of charged AdS black hole \cite{Spallucci,LML}, the general approaches to the critical exponents of van der Waals type phase transition in AdS black holes \cite{Majhi:2016txt,Mandal:2016anc,Bhattacharya:2017hfj,Bhattacharya:2017nru,Bhattacharya:2019qxe}, quasinormal modes behavior near the phase transition points \cite{LZW}, van der Waals behavior of entanglement entropy in charged AdS black hole \cite{ZengLi,LiWei}, hairy black hole chemistry \cite{AMR}, Ruppeiner geometry \cite{Zangeneh:2016fhy,Sahay,WeiLiuprd2019,XWY,GB,Ghosh:2019pwy}, and black hole microstructure \cite{WeiLiuPRL,WeiLiuMann}. One can refer to \cite{reviewPVcriticality} for a comprehensive recent review of P-V criticality of charged AdS black holes.

In thermodynamics and statistical physics, system is considered to be made up of molecules, and the macroscopic properties can be determined by the degrees of freedoms from microscopic molecules. The liquid or gas or solid form of materials is in this sense a macroscopic emergent state. In general, for one certain system considered, there are many possible microscopic states. Each state has a different weight or probability to appear. These emergent states have larger weights which are associated with the thermodynamic free energy by the Boltzmann law. The free energy distributed in the whole state space forms a free energy landscape \cite{Goldenfeld,FSW,FW}. The free energy landscape can be characterized and studied by its dependence on the order parameters such as density in van der Waals liquid-gas system. It should be noted that the temperature can modulate the free energy landscape on the order parameter space. In this picture, the local stationary state, which is represented by the local basin (minimum) of the free energy landscape, can have the probability to switch to the global stationary state represented by the global basin (minimum) and vice versa. Then any state has a life time due to the chance of transition to other states. Notice that the thermal fluctuation is the reason behind the kinetics of this type of  state switching.

In this work, inspired by the analogy of RNAdS black holes and van der Waals liquid system, we will go a few steps further and gain deeper understanding on this by studying the thermodynamics and the underling kinetics of the phase transition between the small and the large RNAdS black holes based on the free energy landscape. The free energy landscape formalism for the black hole system is formulated in the recent paper \cite{LW}. It has been applied to study the thermodynamics and the kinetics of Hawking-Page phase transitions in Einstein gravity and massive gravity \cite{LW}.

The main assumptions to formulate the free energy landscape for black hole system are described as follows \cite{LW}. Black hole, which is considered as a thermal entity, should be a macroscopic emergent state of the underlying degrees of freedom of the spacetime. From the phase transition viewpoint, order parameter becomes very useful and important for describing the degree of freedom at the emergent level. We argue that the radius of the black hole event horizon can be taken as the proper order parameter to formulate the free energy landscape. This point is different but related to the viewpoint that the number density of the black hole molecules is the order parameter to measure the microscopic degrees of freedom \cite{WeiLiuPRL,WeiLiuMann}. In the previous studies on the small/large RNAdS black hole phase transition, there are three branches of black holes (the small, the large, and the intermediate black holes), which are distinguished by their radii of the event horizons. We want to study the phase transition and the kinetics based on the free energy landscape, where the free energy is defined as the continuous function of the order parameter. For these reasons, we further assumed that, besides the small, the large, and the intermediate black holes, there should be transient states during the phase transition process. Therefore, we have assumed that there exists a series of black hole spacetimes with a wide range of radius at the specific temperature. These spacetime states compose the canonical ensemble we are considering. In other words, this ensemble includes the small, the large, and all the transient states during the phase transition. The free energy landscape is formulated by specifying every spacetime state in the ensemble a Gibbs free energy. In the present case, this can be achieved by generalizing the on-shell Gibbs free energy for the three branches of RNAdS black holes to all spacetime states. The generalized Gibbs free energy, which can be expressed as a function of the order parameter and the ensemble temperature, is off-shell because the assumed transient spacetime states are not the solutions to the Einstein equations. Using the Gibbs free energy topography, the emergence of the phases and the associated phase transition can be analyzed. Furthermore, by adjusting the ensemble temperatures, we can also analyze the phase diagram and explore the thermodynamic stability.

For the underling kinetics of black hole phase transition, it is proposed that the stochastic dynamics of black hole phase transition under thermal fluctuations can be studied by using the associated probabilistic Fokker-Planck equation on the free energy landscape \cite{NSM,WangPRE,WangJCP,JW,BW}.
In this regard, due to the thermal fluctuation, a small (large) black hole state can have the chance to switch to a different state such as large (small) black hole state through the thermal free energy barrier crossing process. By solving the Fokker-Planck equation with proper boundary condition, we can obtain the stationary probability distribution of black hole states at different temperatures as well as the mean first passage time of the kinetics process from the small black hole to the large black hole and vice versa. On the free energy landscape, kinetics are closely related to the shape of the free energy landscape when varying the temperatures \cite{WangPRE,WangJCP,JW,BW}. We further study the statistical fluctuations and distributions of the kinetics at various temperatures. Again the landscape topography is important for characterizing the behaviors of the fluctuations in kinetics.

In principle, the free energy landscape formulism can be applied to study any kind of phase transition of black hole system. In \cite{LW}, it has been applied to study the Hawking-Page phase transitions in Einstein gravity and in massive gravity. This provides us significant insights into understanding the underling thermodynamics and kinetics of the Hawking-Page phase transition. In the present work, our aim is to apply this formulism to study the van der Waals type phase transition in RNAdS black holes. This type of phase transition is very different from the Hawking-Page transition and of very important significance in studying the microstructures of AdS black holes \cite{Wei:2019yvs,Wei:2019ctz,Ghosh:2020kba,Yerra:2020oph,Dehyadegari:2020ebz,Sheykhi:2019vzb}. It should be noted that our method in this work is also different from that used in \cite{LW}. In this work, we study the kinetics of the first passage process by numerically solving the Fokker-Planck equation. In this way, we can obtain not only the mean first passage time and its statistical fluctuation but also the probabilistic evolution of spacetime state and the distribution of the first passage time. In \cite{LW}, the analytical expressions for the mean first passage time and its fluctuation were derived, from which the kinetics of the first passage process is studied. In this sense, this work provides an alternative method to study the kinetics and its distribution of the phase transition in the black hole system.

This paper is arranged as follows. In section II, we explore the thermodynamics and phase transition of RNAdS black hole through free energy landscape. In section III, we study the Fokker-Planck equation on the free energy landscape for quantifying the probabilistic evolution of the underlying stochastic dynamics. The time dependent probabilistic evolution solutions are also obtained by numerically solving Fokker-Planck equation with the reflecting boundary conditions. In section IV, we present the numerical results of the distributions of first passage time switching from small (large) to large (small) black hole phases and the corresponding mean first passage time and its fluctuations at different temperatures. The conclusion and discussion are presented in the last section.

\section{Thermodynamics of RNAdS black hole}

\subsection{Thermodynamic characterization}

We start with the metric of RNAdS black hole, which describes the spherically symmetric charged black hole solution to Einstein-Maxwell action with negative cosmological constant \cite{Carter}. The line element is given by ($G=1$ units)
\begin{eqnarray}\label{metric}
ds^2=-\left(1-\frac{2M}{r}+\frac{Q^2}{r^2}+\frac{r^2}{L^2}\right)dt^2
+\left(1-\frac{2M}{r}+\frac{Q^2}{r^2}+\frac{r^2}{L^2}\right)^{-1}dr^2
+r^2d\Omega^2\;,
\end{eqnarray}
where $M$ is the mass, $Q$ is the charge, and $L=\sqrt{\frac{-3}{\Lambda}}$ is the AdS curvature radius with $\Lambda$ being the cosmological constant.

The event horizon is determined by the largest root of the equation
\begin{eqnarray}\label{metricfunction}
f(r)=1-\frac{2M}{r}+\frac{Q^2}{r^2}+\frac{r^2}{L^2}=0\;.
\end{eqnarray}
The mass of black hole can be expressed in terms of the horizon radius $r_+$ as
\begin{eqnarray}\label{Mass}
M=\frac{r_+}{2}\left(1+\frac{r_+^2}{L^2}+\frac{Q^2}{r_+^2}\right)\;.
\end{eqnarray}
The Hawking temperature is given by
\begin{eqnarray}\label{Hawkingtemperature}
T_H=\frac{1}{4\pi}f'(r)|_{r=r_+}=\frac{1}{4\pi r_+}\left(1+\frac{3r_+^2}{L^2}-\frac{Q^2}{r_+^2}\right)\;.
\end{eqnarray}

In the case of an asymptotically AdS black hole in four
dimensions, one can relate the thermodynamic pressure to the cosmological constant as suggested in \cite{Kastor:2009wy,Dolan1,Dolan2}
\begin{eqnarray}\label{Pressure}
P=\frac{3}{8\pi} \frac{1}{L^2}\;.
\end{eqnarray}
Replacing the cosmological constant by thermodynamic pressure in Eq.(\ref{Hawkingtemperature}), Hawking temperature can be rewritten as
\begin{eqnarray}\label{stateequation}
T_H=\frac{1}{4\pi r_+}\left(1+8\pi P r_+^2-\frac{Q^2}{r_+^2}\right)\;.
\end{eqnarray}

\begin{figure}
  \centering
  \includegraphics[width=6cm]{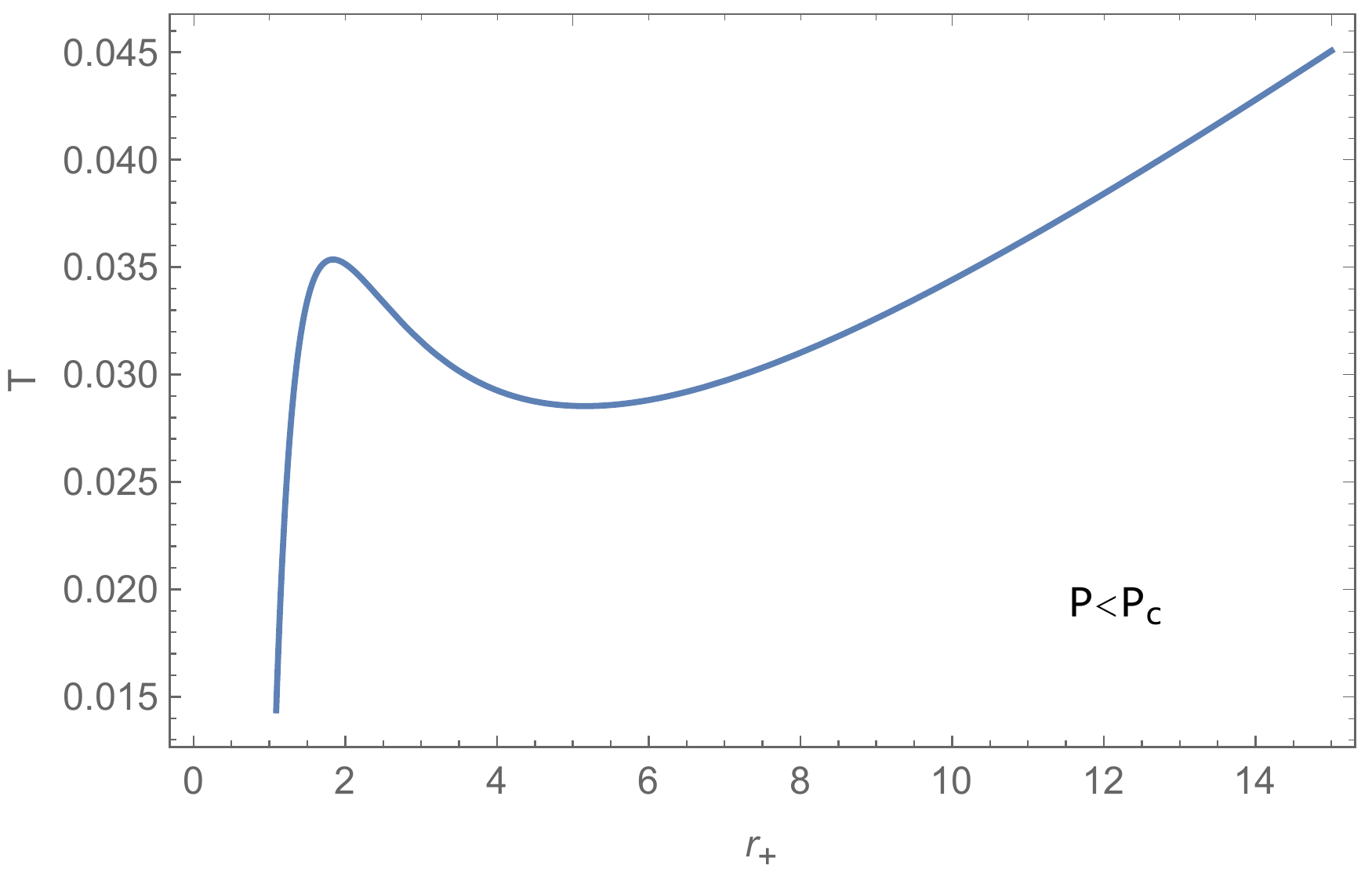}
  \includegraphics[width=6cm]{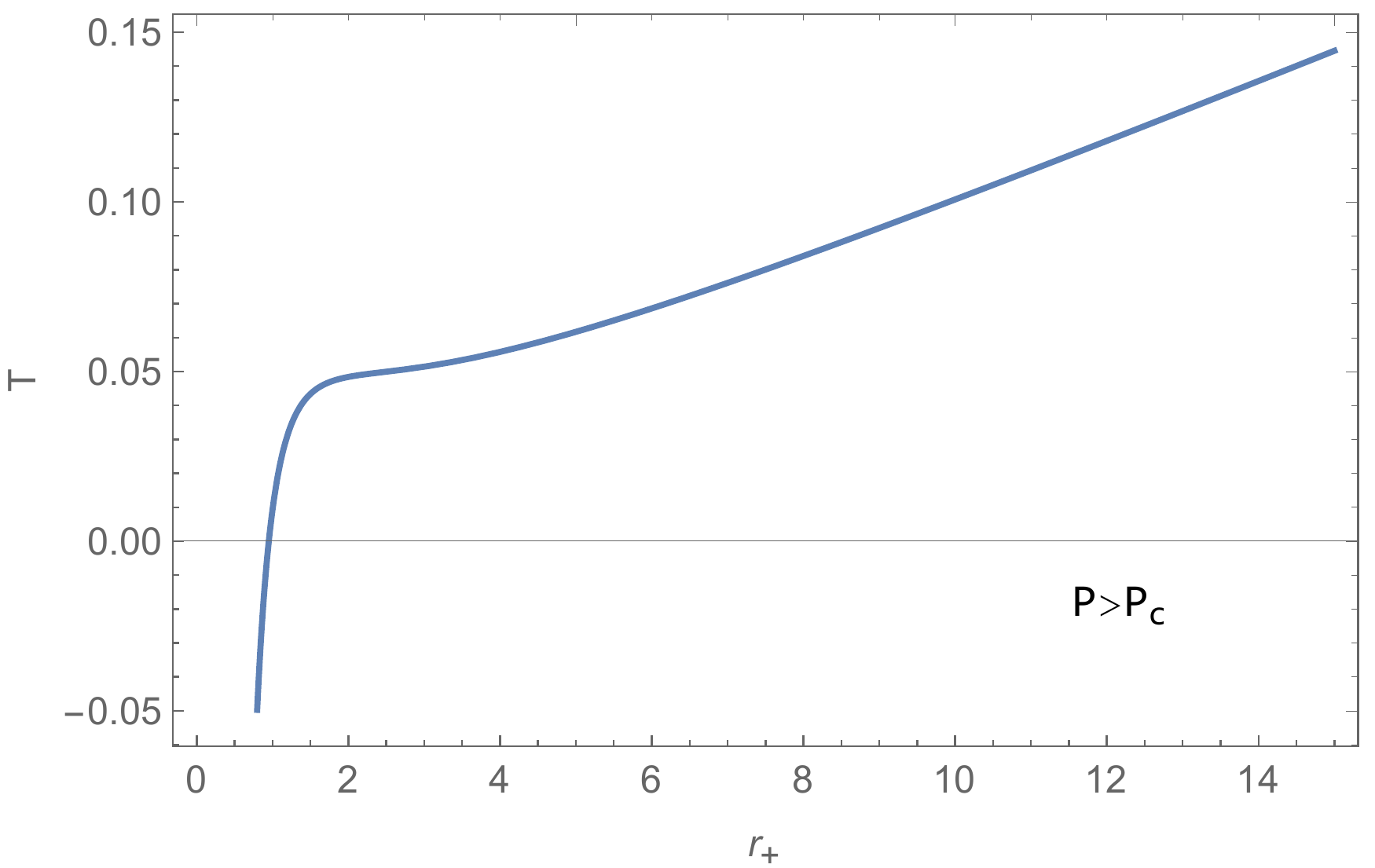}
  \caption{Black hole Temperature $T_H$ as a function of event horizon radius $r_+$ when $P<P_c$ (left pannel) and $P>P_c$ (right pannel).}
  \label{temp}
\end{figure}

There exits a critical pressure $P_c=\frac{1}{96\pi Q^2}$, above which black hole temperature $T_H$ is a monotonic function of black hole radius $r_+$, while below which black hole temperature $T_H$ have the local minimal and local maximum value. In Fig.\ref{temp}, we have depicted the black hole temperature as a function of black hole radius $r_+$ when $P<P_c$ (left pannel) and $P>P_c$ (right pannel).

When $P<P_c$, the local minimal and local maximum values of black hole temperature are determined by equation
\begin{eqnarray}
\frac{\partial T_H}{\partial r_+}=0\;,
\end{eqnarray}
which gives us the solutions
\begin{eqnarray}
r_{min/max}=\left[\frac{1\pm\sqrt{1-96\pi P Q^2}}{16\pi P}\right]^{1/2}\;.
\end{eqnarray}
when substituting back to Eq.(\ref{stateequation}), one can get the local minimal and local maximum values of black hole temperature which are respectively given by
\begin{eqnarray}\label{Tminmax}
T_{min/max}&=&\frac{2\sqrt{P}}{\sqrt{\pi}}\frac{\left(1-32\pi P Q^2\pm\sqrt{1-96\pi P Q^2}\right)}{\left(1\pm\sqrt{1-96\pi P Q^2}\right)^{3/2}}.
\end{eqnarray}

When $T_{min}<T_H<T_{max}$, there exists three branches of black hole solution (i.e. small, intermediate, and large black hole). The intermediate solution is unstable. Similar to the van der Waals liquid-gas system, there is a first order phase transition from the small black hole to the large black hole.

\subsection{Free energy landscape}

As discussed in the Introduction, we consider the canonical ensemble at the specific temperature $T$ which is composed by a series of black hole spacetimes with an arbitrary horizon radius.
Note that there are three branches of the black holes (the small, the large,and the intermediate black holes) are the on-shell solutions to the stationary Einstein field equations. Other spacetimes are off-shell and do not satisfy the stationary Einstein field equation. The small and the large black hole states are locally or globally stable. Other spacetime states are the unstable transient or excited states, which are bridges for the state switching or phase transition processes. The on-shell Gibbs free energy for the three branches of RNAdS black hole can be given by the thermodynamic relationship $G=M-T_H S$ or calculated directly from the Euclidean action \cite{Dolan2}.
Replacing the Hawking temperature $T_H$ by the ensemble temperature $T$ in the on-shell Gibbs free energy expression, we can define the generalized off-shell Gibbs free energy for the transient black hole state, which is given by
\begin{eqnarray}\label{GibbsEq}
G=M-TS=\frac{r_+}{2}\left(1+\frac{8}{3}\pi P r_+^2+\frac{Q^2}{r_+^2} \right)-\pi T r_+^2\;.
\end{eqnarray}

\begin{figure}
  \centering
  \includegraphics[width=6cm]{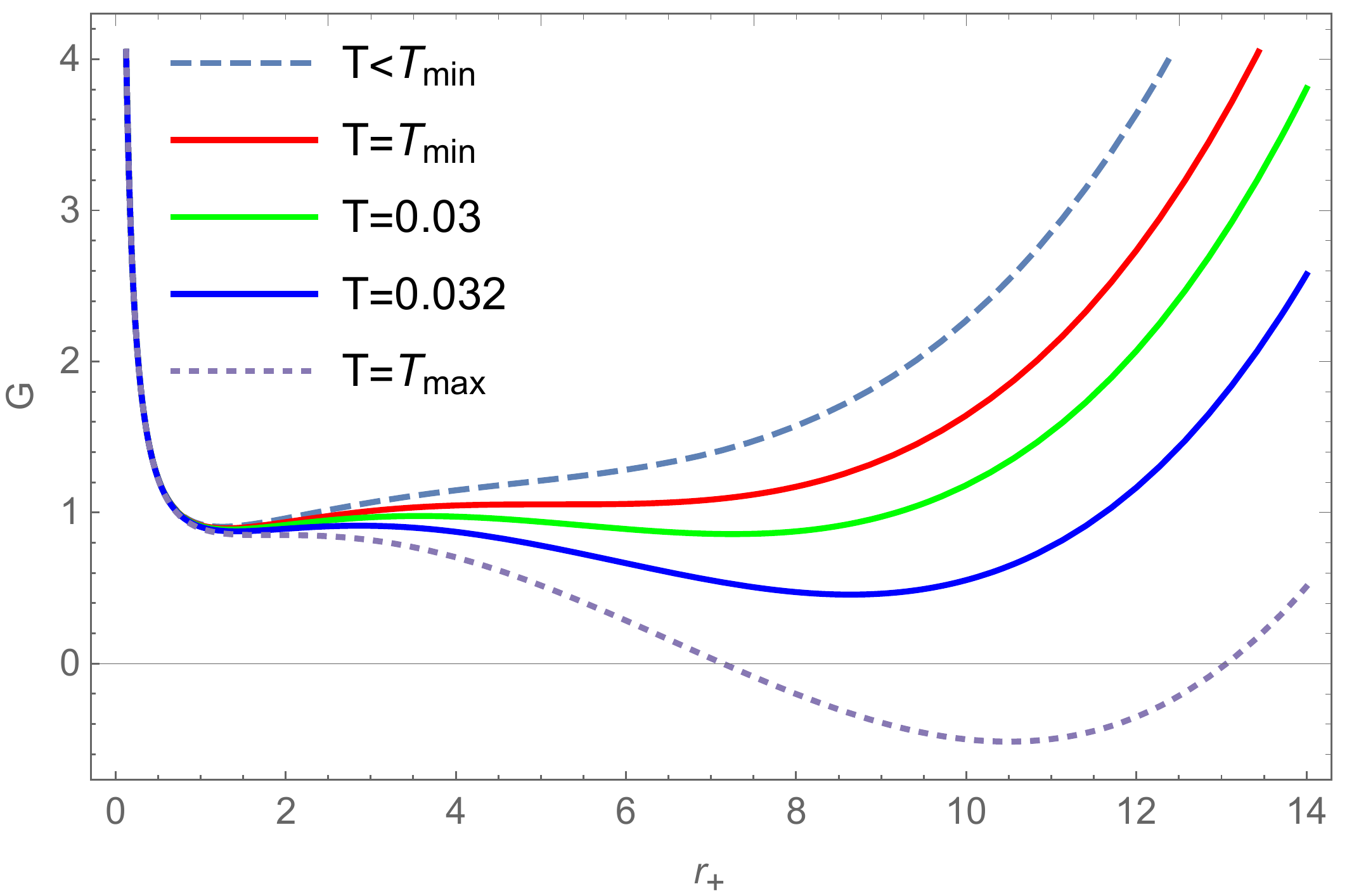}\\
  \caption{Gibbs free energy as a function of $r_+$ for $P<P_c$ with different temperatures. }
  \label{Gibbs}
\end{figure}

This Gibbs free energy should be considered as the generalized canonical free energy, where the horizon radius $r_+$ can take all values from zero to infinity because $r_+$ is taken as the order parameter to describe the microscopic degrees of freedom \cite{LW}. This type of the generalized free energy has been applied to investigate and understand the physical process of the phase transitions in the Schwarzschild black holes \cite{York,AL}. In this way, we can then formulate the free energy landscape for the RNAdS black holes. Thus we can quantify the free energy landscape by plotting the free energy as a function of black hole radius $r_+$ as shown in Fig. \ref{Gibbs} for different temperatures. In the plot, $P=0.4 P_c$, $Q=1$, $T_{min}=0.0285$, and $T_{max}=0.0354$.

It can be seen when $T_{min}<T<T_{max}$ the Gibbs free energy has double basin (well) shape. In this case, the Gibbs free energy has three local extremals (two stable and one unstable), which are determined by the equation
\begin{eqnarray}\label{pgpr}
\frac{\partial G}{\partial r_+}=\frac{1}{2}+4\pi P r_+^2-\frac{Q^2}{2r_+^2}-2\pi T r_+=0\;.
\end{eqnarray}
This is just the equation for the black hole temperature $T_H$. By solving this equation for $r_+$, one can get the lengthy expressions of the radii for the small, large, and intermediate black holes. We denote them as $r_s, r_m$, and $r_l$, respectively, which are determined by $T, P$, and $Q$. The intermediate black hole which corresponds to the maximum value of the Gibbs free energy is unstable. The small and large black holes are all locally stable, and the thermodynamically stable black hole has the minimal Gibbs free energy.

\begin{figure}
  \centering
  \includegraphics[width=6cm]{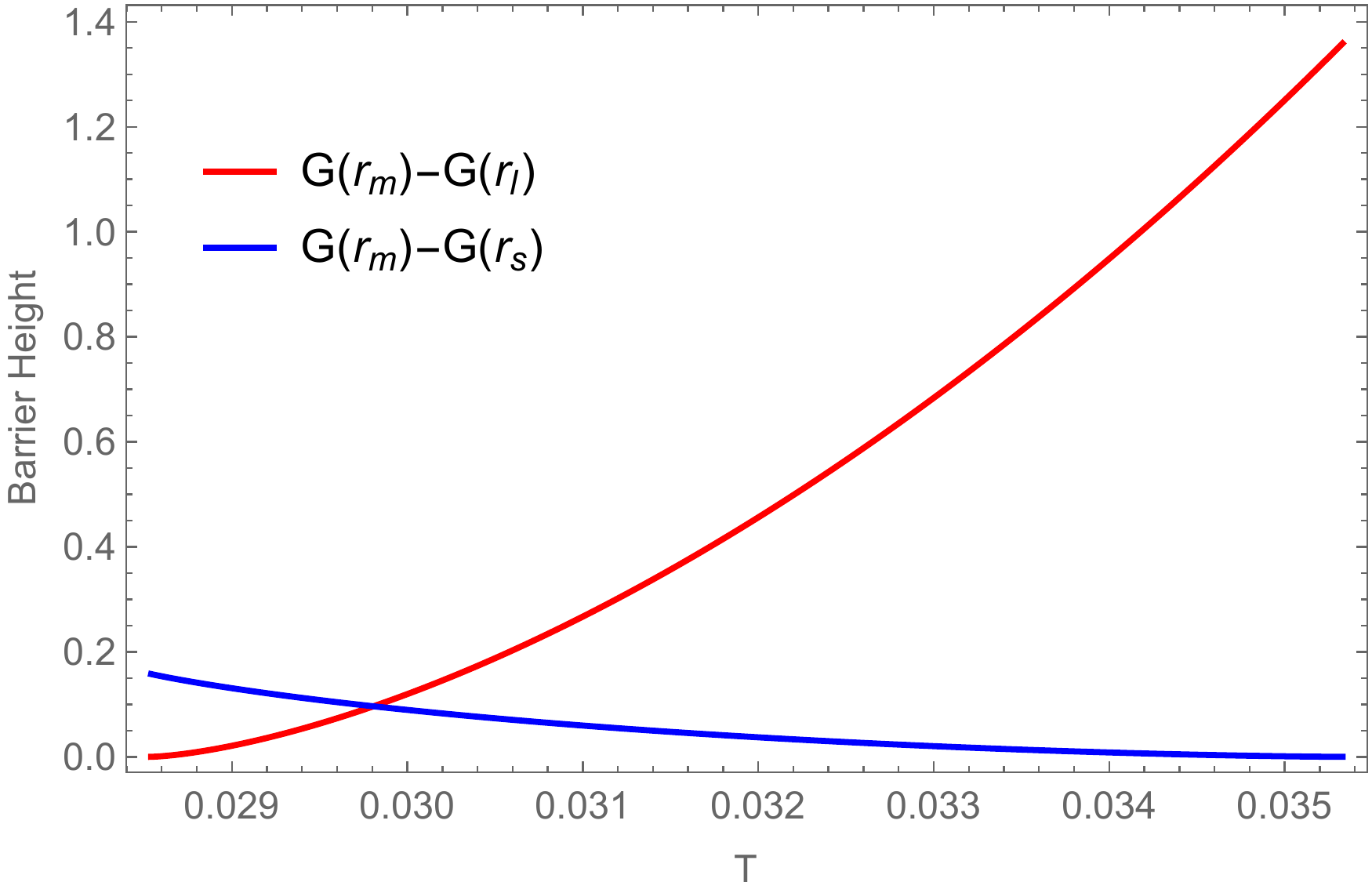}\\
  \caption{Barrier height as a function of temperature $T$. The red line represents the barrier height between the intermediate black hole and the large black hole, while the blue line represents the barrier height between the intermediate black hole and the small black hole.}
  \label{Barrierheight}
\end{figure}

Gibbs free energy for the the small, large, and intermediate black holes are given by
\begin{eqnarray}
G_{s/m/l}=\frac{r_{s/m/l}}{4}-\frac{2}{3}\pi P r_{s/m/l}^3+\frac{3}{4} \frac{Q^2}{r_{s/m/l}}\;.
\end{eqnarray}
The free energy landscape topography quantified by the free energy barrier heights from the small black hole to the large black hole and from the large black hole to the small black hole are plotted in Fig.\ref{Barrierheight}. It is obvious that they are all monotonic functions of the black hole temperature. When temperature increases, the free energy barrier from small black hole state to the large black hole state decreases while the free energy barrier from large black hole state to the small black hole state increases. This is consistent with the free energy landscape as a function of black hole radius at varying temperatures as shown in Fig.\ref{Gibbs}.

\subsection{Thermodynamic stability of emergent phases, phase transitions and phase diagram}

There exists a transition temperature determined by the Gibbs free energies of the large black hole and small black hole are equal. The transition temperature should be determined by the following equations
\begin{eqnarray}
&&\frac{1}{2}+4\pi P r_s^2-\frac{Q^2}{2r_s^2}-2\pi T r_s=0\;,\nonumber\\
&&\frac{1}{2}+4\pi P r_l^2-\frac{Q^2}{2r_l^2}-2\pi T r_l=0\;,\nonumber\\
&&\frac{r_{s}}{4}-\frac{2}{3}\pi P r_{s}^3+\frac{3}{4} \frac{Q^2}{r_{s}}=
\frac{r_{l}}{4}-\frac{2}{3}\pi P r_{l}^3+\frac{3}{4} \frac{Q^2}{r_{l}}\;.
\end{eqnarray}

\begin{figure}
  \centering
  \includegraphics[width=6cm]{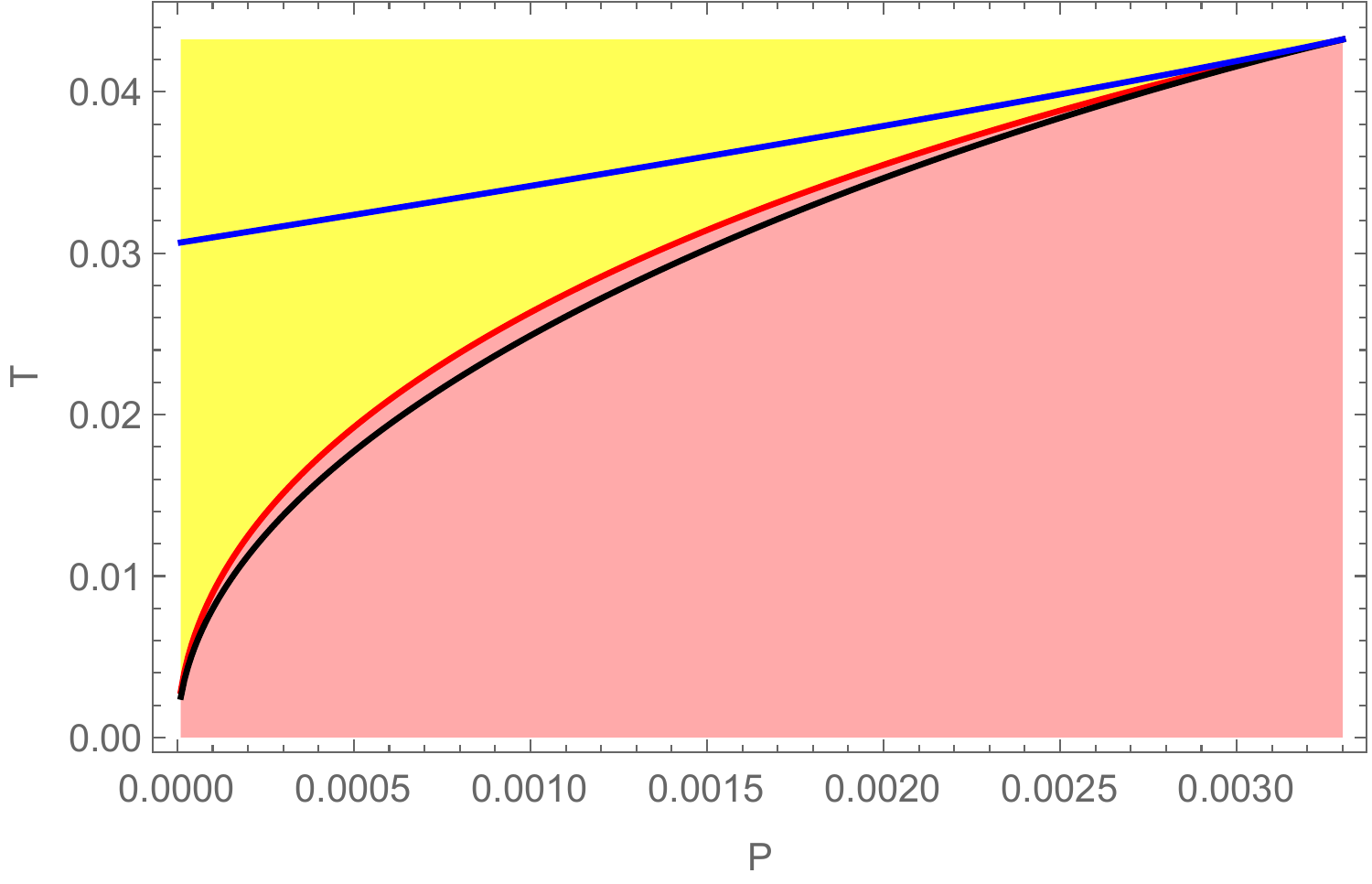}\\
  \caption{Phase diagram of RNAdS black holes. The blue, black, and red lines are the plots of $T_{max}$, $T_{min}$, and $T_{trans}$ as a function of pressure $P$. In these plots, the range of $P$ is from $0$ to the critical pressure $P_c$. The small and large black holes are thermodynamically stable in pink and yellow regions, respectively. }
  \label{phasediagram}
\end{figure}

In Fig.\ref{phasediagram}, we provided the thermodynamic phase diagram by plotting the transition temperature as well as $T_{min/max}$ as a function of pressure $P$. These curves separate the $P-T$ plane into four thermodynamic phase regions. In the phase regions of below the black line and above the blue line, there is only one black hole solution, and it is always thermodynamic stable. In the middle phase region, there are three black hole solutions as discussed above. The red line represents the transition temperature, where the free energies for the small black hole and the large black hole are equal, i.e. the red line represents the coexistence curve that the small and the large black hole phases coexists in equal probability. In the phase region between the red line and the black line, the free energy of the small black hole is less than the free energy of the large black hole, while in the phase region between the red line and the blue line this relation is reversed. Therefore, we can conclude that the small black hole is thermodynamically stable in the phase region between the red line and the black line and the large black hole is thermodynamically stable in the phase region between the red line and the blue line. However, from the ensemble viewpoint, the stable black hole states still have the chance of switching into other black hole state due to the presence of the thermal fluctuations. We will examine this by studying the corresponding kinetics process described by the probabilistic Fokker-Planck equation in the following sections.

The secondary derivative of Gibbs free energy is given by
\begin{eqnarray}
\frac{\partial^2 G}{\partial r_+^2}=8\pi P r_+ +\frac{Q^2}{r_+^3}-2\pi T\;.
\end{eqnarray}
By solving the equation $\frac{\partial^2 G}{\partial r_+^2}=0$ together with Eq.(\ref{pgpr}), one can also determine the local minimal and local maximum values of temperature which are given by Eq.(\ref{Tminmax}).

\section{Fokker-Planck equation governing the probabilistic evolution on the free energy landscape}

In this section, we will discuss the transition state theory of black hole state switching or phase transition by treating black hole as a state in extended phase space. As shown in the last section, we know that Gibbs fee energy landscape as a function of black hole radius exhibits double basin (well) shape when $T_{min}<T<T_{max}$. The two local minima correspond to the small and the large black holes respectively. One naturally regards the black hole radius as the reaction coordinate or order parameter.

From now on, we use the symbol $r$ to denote the black hole radius $r_+$ for the sake of simplicity. In the formalism of free energy landscape, the Gibbs free energy $G(r)$ is a function of the order parameter $r$, where $r$ denotes the black hole radius. Our aim is to study the evolution of the system under the thermal fluctuation. The probability distribution of these states evolving in time should be a function of the order parameter $r$ (black hole radius) and time $t$. Thus, the
probability distribution of spacetime state in the ensemble is denoted by $\rho(r, t)$. So we firstly write down the corresponding Fokker-Planck probabilistic evolution equation.

The Fokker-Planck equation for the probabilistic evolution on the free energy landscape is explicitly given by \cite{NSM,WangPRE,WangJCP,JW,BW}
\begin{eqnarray}\label{FPequation}
\frac{\partial \rho(r,t)}{\partial t}=D \frac{\partial}{\partial r}\left\{
e^{-\beta G(r)}\frac{\partial}{\partial r}\left[e^{\beta G(r)}\rho(r,t)\right]
\right\}\;.
\end{eqnarray}
In the above equation, the inverse temperature is $\beta=1/kT$ and the diffusion coefficient is  $D=kT/\zeta$ with $k$ being the Boltzman constant and $\zeta$ being dissipation coefficient. Without loss of generality, we will take $k=\zeta=1$ in the following.

In order to solve the Fokker-Planck equation, two types of boundary conditions should be imposed at the boundaries of computational domain depending on the question we consider. We list the boundary conditions at $r=r_0$ for example.
\begin{itemize}
\item Reflecting boundary condition:
\begin{eqnarray}\label{bc1}
\left.
e^{-\beta G(r)}\frac{\partial}{\partial r}\left[e^{\beta G(r)}\rho(r,t)\right]\right|_{r=r_0}=0\;.
\end{eqnarray}
It is equivalent to
\begin{eqnarray}\label{bc11}
\left.\beta G'(r)\rho(r, t)+\rho'(r, t)\right|_{r=r_0}=0\;.
\end{eqnarray}

\item Absorbing boundary condition:
\begin{eqnarray}\label{bc2}
\rho(r_0,t)=0\;.
\end{eqnarray}
\end{itemize}

In this section, we study the time evolution of probability of state distribution in the canonical ensemble, which is composed by a series of black hole spacetime with the radius ranging from $0$ to infinity. It is obvious that the Gibbs free energy in Eq. (\ref{GibbsEq}) is divergent at $r=0$ and $r=+\infty$. This means that the probability should be preserved in the time evolution process because the Gibbs free energy can be considered as the effective potential. The reflecting boundary condition will preserve the normalization of probability distribution or the probability conservation (the total probability as the sum of all should be equal to $1$). In practice numerical computation, we set the reflecting boundary condition at $r=0.1$ and $r=12$ in order to avoid the numerical instability. We choose the initial condition as
\begin{eqnarray}\label{initial}
\rho(r,0)=\frac{1}{\sqrt{\pi}a} e^{-(r-r_i)^2/a^2}\;,
\end{eqnarray}
with the parameter $a=0.1$. This Gaussian wave packet is a good approximation of $\delta$-distribution in numerical computation process. It should be noted that the initial wave pocket is normalized. We have checked that the normalization of probability distribution is preserved in the evolution process. We take $r_i=r_s$ or $r_i=r_l$ as the initial condition of system, which means that the system initially is either in small or large black hole state.

\begin{figure}
  \centering
  \includegraphics[width=12cm]{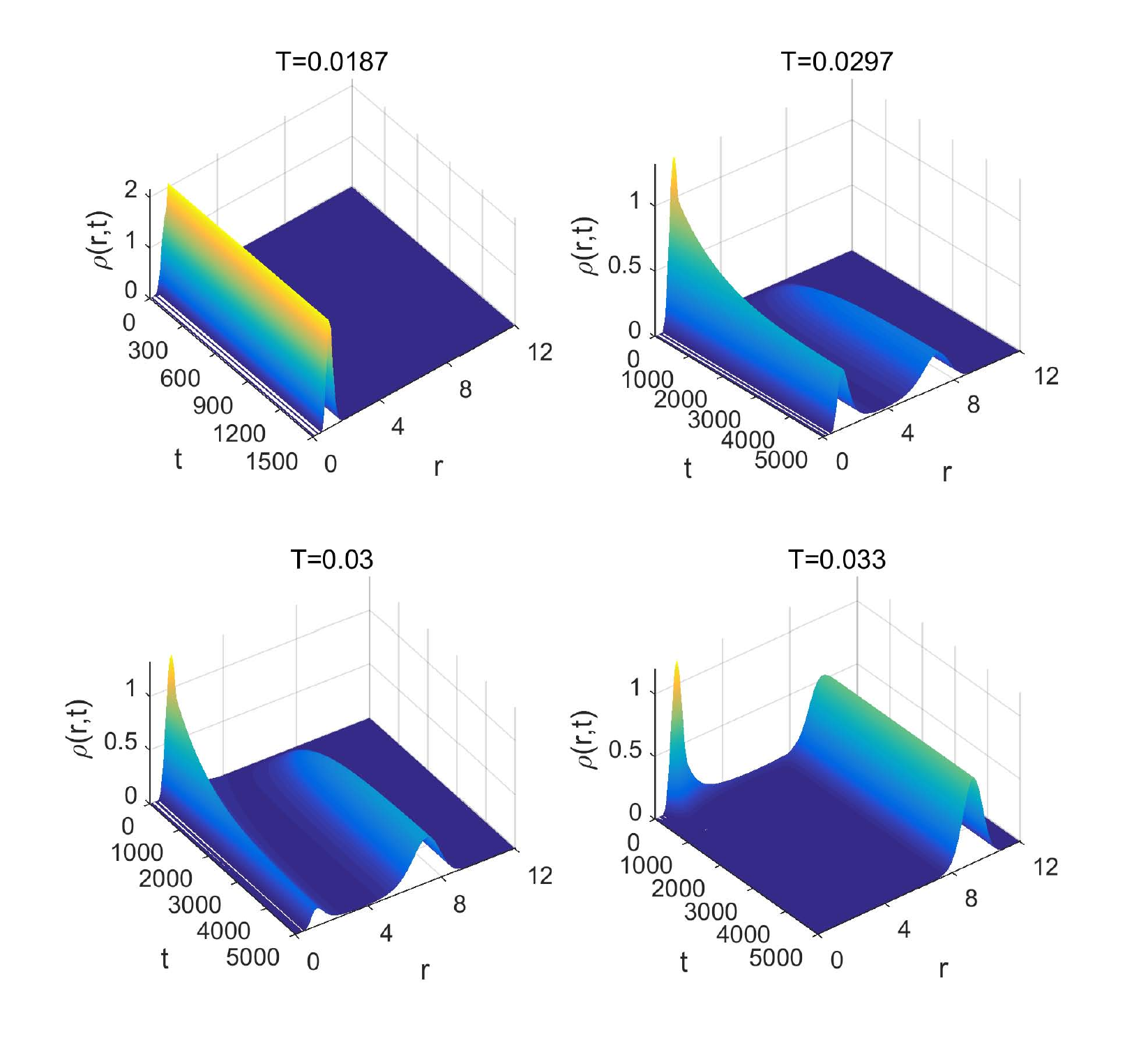}
  \caption{The distributions of probability $\rho(r, t)$ at different temperatures
  $T=0.0187, 0.0297, 0.03$ and $0.033$. The initial wave packet is located at the small black hole representation.}
  \label{FPsolutionstl}
\end{figure}

\begin{figure}
  \centering
  \includegraphics[width=12cm]{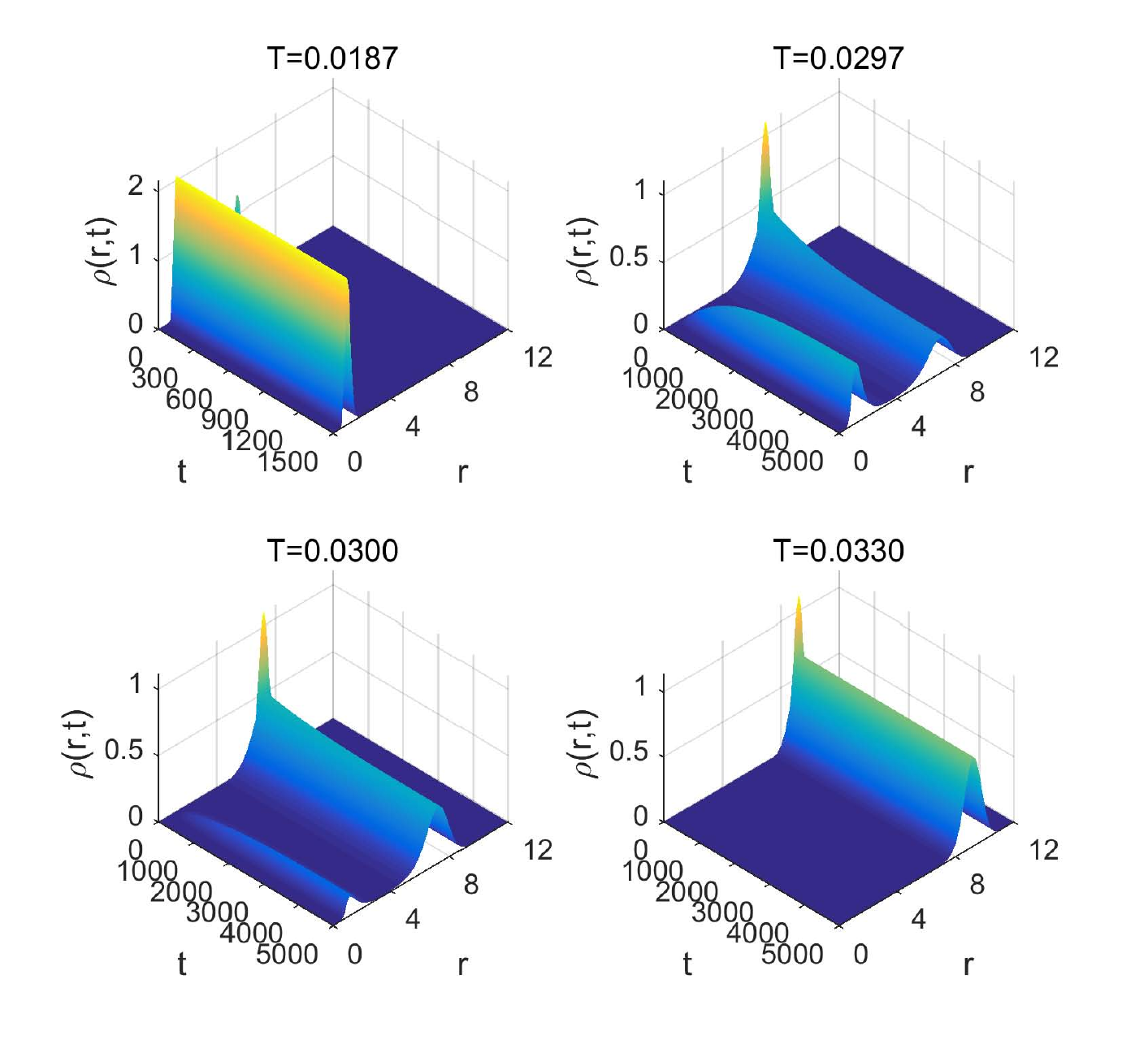}
  \caption{The distributions of probability $\rho(r, t)$ at different temperatures
  $T=0.0187, 0.0297, 0.03$ and $0.033$. The initial wave packet is located at the large black hole representation.}
  \label{FPsolutionlts}
\end{figure}

The time dependent behaviors of the probability distribution of the black holes in extended phase space at different temperatures are plotted in Fig.\ref{FPsolutionstl} and Fig.\ref{FPsolutionlts}.
In Fig.\ref{FPsolutionstl} and Fig.\ref{FPsolutionlts}, the initial probability distributions are Gaussian wave pockets located at the small and the large black hole radii respectively.
In these plots, it can be observed that, the probability distribution of the black hole states reaches a quasi-stationary distribution very quickly at very early time, and then the final stationary distribution will be saturated at long time limit. It is shown the peak shape of the initial distribution will become smooth distribution within the free energy basin ( well) where the initial wave pocket locates. This quasi-stationary distributions of probability at very early time are only determined by the local free energy basin (well), and are not affected by the existence of the other free energy basin (well). Then the smooth quasi-stationary distributions will spread out to the other free energy basin (well) along the time.

The final stationary distributions are determined by $\rho_{st}(r)\propto e^{-G(r)/T}$, which can be obtained easily from the Fokker-Planck equation by setting $\frac{\partial\rho(r, t)}{\partial t}=0$. This is consistent with the Boltzmann relationship between the free energy and the equilibrium probability distribution. Therefore, the stationary distribution from the long time evolution reaches the equilibrium probability. The thermodynamic stable state is then determined by the maximum of the final stationary distribution. However, on the free energy landscape theory, black hole state can have the chance of escape from one state to another due to the presence of thermal fluctuations.

Since the Gibbs free energy is inversely related to the probability by the Boltzman law in equilibrium thermodynamics, the large free energy corresponds to the low probability or the weight of the black hole spacetime states. In our case, the Gibbs free energy has a double basin shape. When the order parameter is close to zero or infinity, the free energy becomes infinite and the probability becomes zero. This means the chance or the weight of the black hole spacetime appearance is zero in these extremely large or small regimes of the order parameter. Therefore, our system never reaches very low or very high order parameter regimes and is therefore in practice confined in a finite regime of the order parameter from certain finite small value to certain finite large value. This point is relevant to our numerical calculations of the probability evolution, where the domain of the order parameter is set to be a finite interval in order to avoid the numerical instability.

\section{Kinetics and its fluctuations of black hole state switching dynamics}

In this section, we will study the kinetics by first passage process from one black hole state to another black hole state on the underlying free energy landscape. First passage time is a very important quantity in transition state theory, which can be defined as the time required for the state of the black hole to reach the intermediate transition state (represented by free energy barrier top) of the black hole for the first time in the present case. The mean first passage time defines an average timescale for a stochastic event to first occur.

Firstly, let us consider the first passage time of system from the small black hole state to the large black hole state. We denote the distribution of first passage times by $F_p(t)$ and define $\Sigma(t)$ to be the probability that the state of black hole has not made a first passage by time $t$. The distributions $F_p(t)$ and $\Sigma(t)$ are related by
\begin{eqnarray}\label{FPTequation}
F_p(t)=-\frac{d\Sigma(t)}{dt}\;.
\end{eqnarray}
According to the definition, $\Sigma(t)$ is defined as the probability of black hole
being in the system at time $t$. So we have
\begin{eqnarray}\label{Sigmaequation}
\Sigma(t)=\int_{0}^{r_m} \rho(r, t) dr\;.
\end{eqnarray}

It is obvious that $F_p(t)dt$ is the probability that a small black hole state passes through the intermediate transition state of the black hole (free energy barrier top) for the first time in the time interval $(t, t+dt)$. In this setup, we have made the assumption that the time taken from the intermediate transition state of the black hole to the large black hole state is much smaller than the first passage time. Suppose there is a perfect absorber placed at the site $r_m$ (transition state at the free energy barrier top). If the state of a black hole makes the first passage under the thermal fluctuation, this black hole state leaves the system. The normalization of the probability distribution will not be preserved in this case. Therefore, at very late time, the probability of the black hole still in the system becomes zero, i.e. $\Sigma(r, t)|_{t\rightarrow +\infty}=0$.

By substituting Eq.(\ref{Sigmaequation}) into eq.(\ref{FPTequation}), and using the Fokker-Planck equation (\ref{FPequation}), one can get
\begin{eqnarray}\label{FPTcal}
F_p(t)&=&-\frac{d}{dt}\int_{0}^{r_m} \rho(r, t) dr\nonumber\\
&=&-\int_{0}^{r_m}\frac{\partial}{\partial t} \rho(r, t) dr\nonumber\\
&=&-D\int_{0}^{r_m}\frac{\partial}{\partial r}\left\{
e^{-\beta G(r)}\frac{\partial}{\partial r}\left[e^{\beta G(r)}\rho(r,t)\right]\right\}  dr\nonumber\\
&=&\left.-D e^{-\beta G(r)}\frac{\partial}{\partial r}\left[e^{\beta G(r)}\rho(r,t)\right]\right|_{0}^{r_m}\nonumber\\
&=&\left.-D\frac{\partial}{\partial r}\rho(r,t)\right|_{r=r_m}\;.
\end{eqnarray}
Note that the reflecting boundary condition is imposed at $r=0$ because the probability will not leak out there, and the absorbing boundary condition is imposed at $r=r_m$ (transition state at free energy barrier top) where the perfect absorber is placed. As mentioned above, the probability is not preserved. The purpose of choosing the absorbing boundary here is to facilitate the first passage time calculation. By solving Fokker-Planck equation with the initial condition and boundary conditions, we can get the time distributions of first passage times. It is usually difficult to solve the combined initial value and boundary value problem described by Eqs.(\ref{FPequation})-(\ref{initial}). So we also invoke the numerical method. Note that, because the Gibbs free energy is divergent at $r=0$, the reflecting boundary condition is imposed at $r=0.1$ here.

\begin{figure}
  \centering
  \includegraphics[width=6cm]{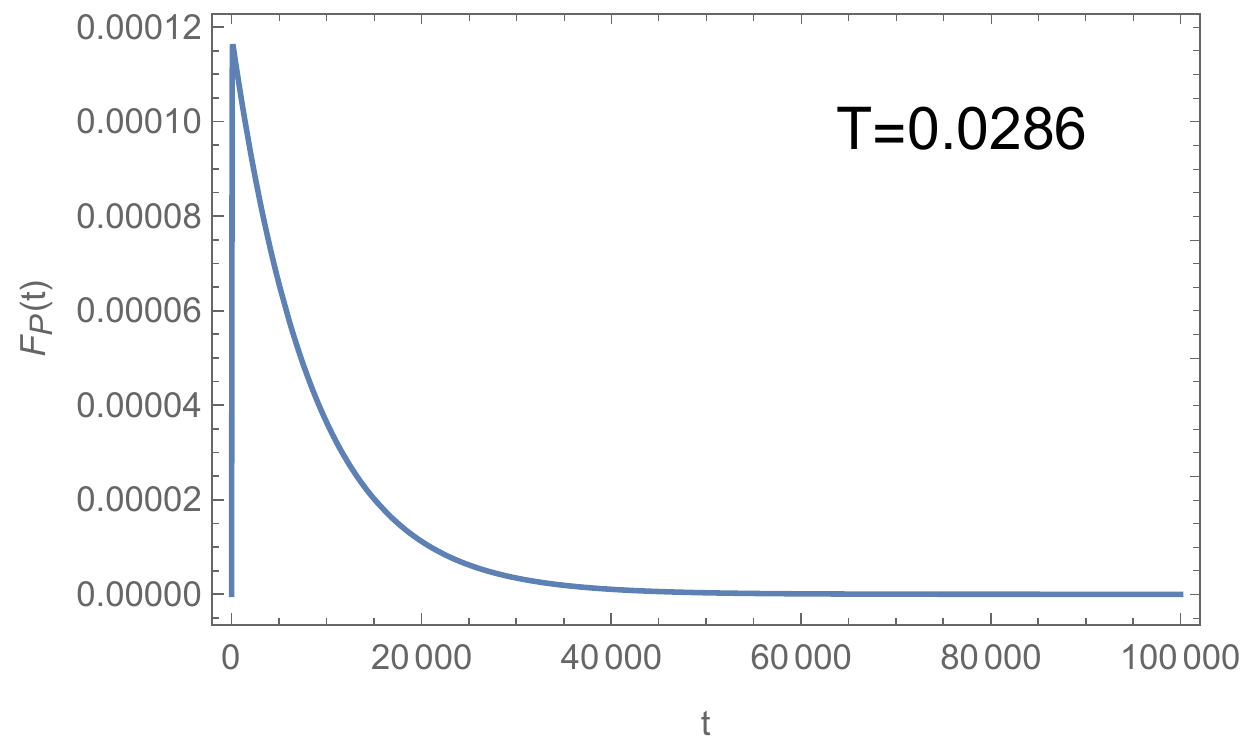}
  \includegraphics[width=6cm]{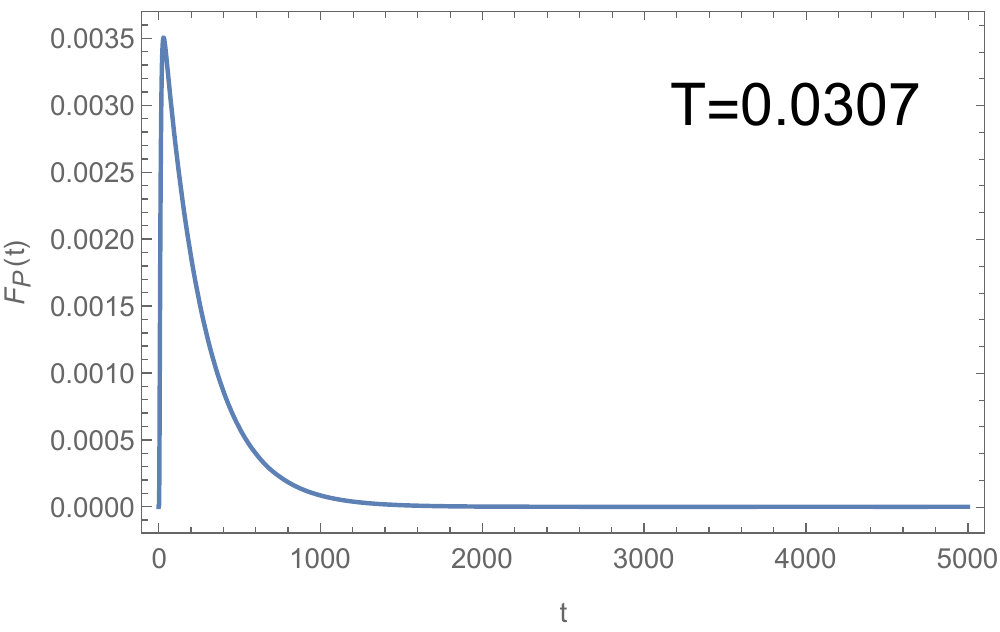}\\
  \includegraphics[width=6cm]{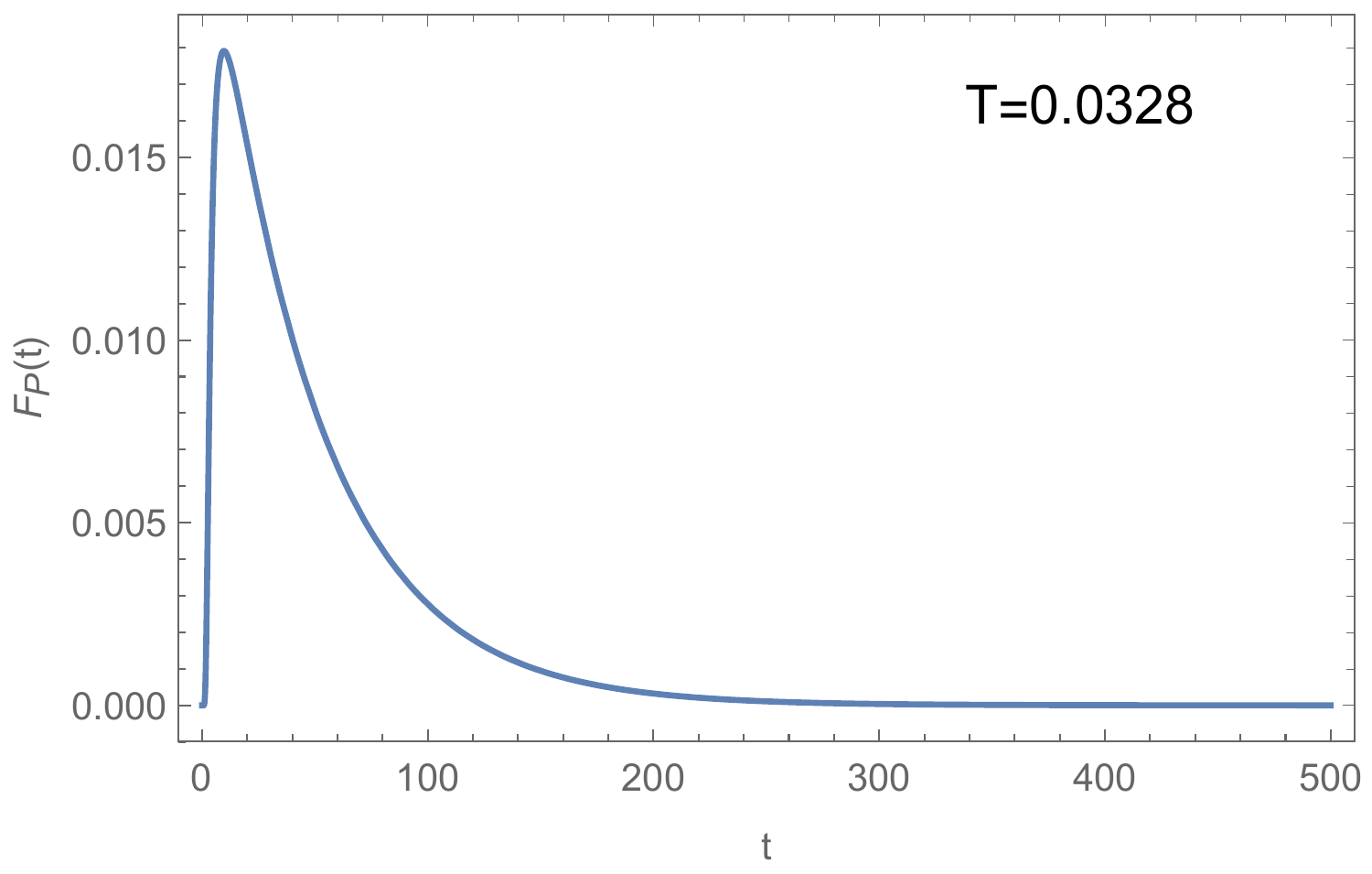}
  \includegraphics[width=6cm]{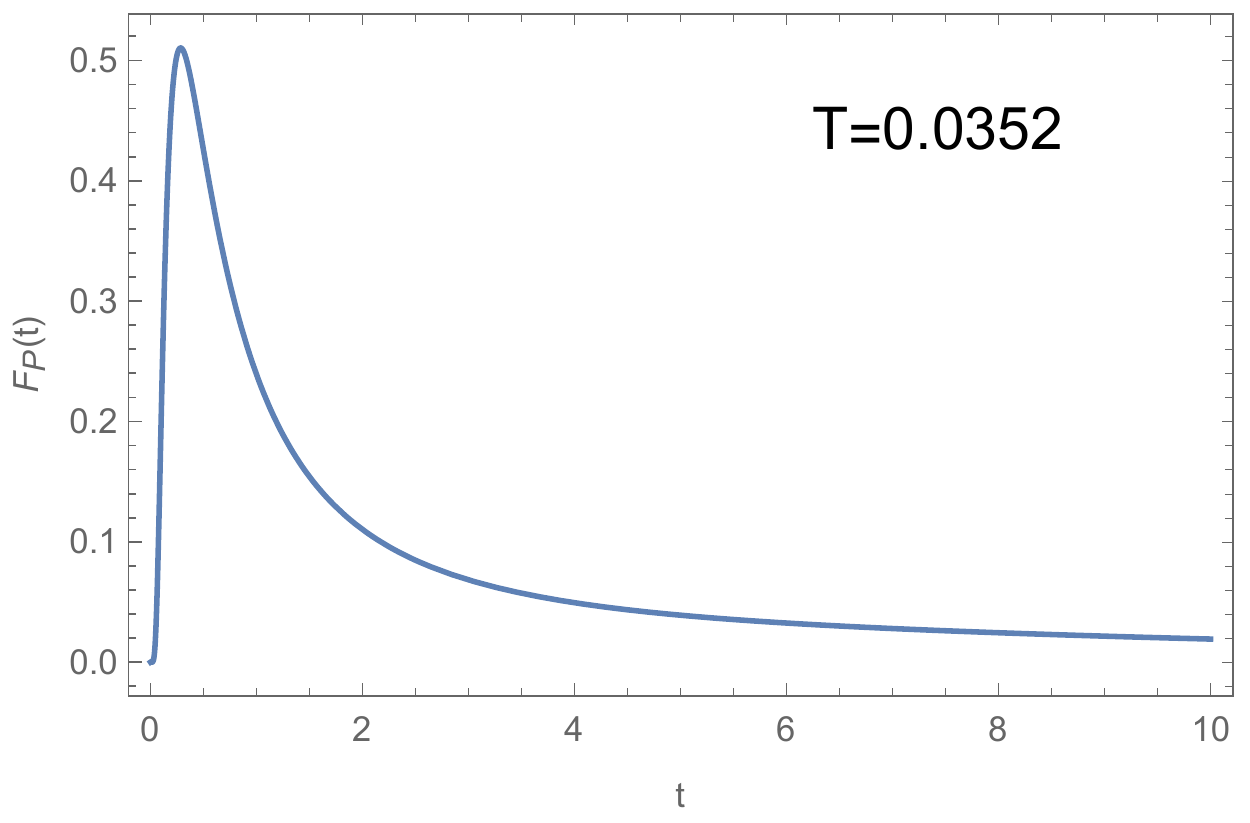}
  \caption{
  The distributions of first passage time $F_p(t)$ from small to large black hole state transition at different temperatures. The initial distribution is Gaussian wave pocket located at the small black hole state.}
  \label{FPTdistributionstl}
\end{figure}

\begin{figure}
  \centering
  \includegraphics[width=6cm]{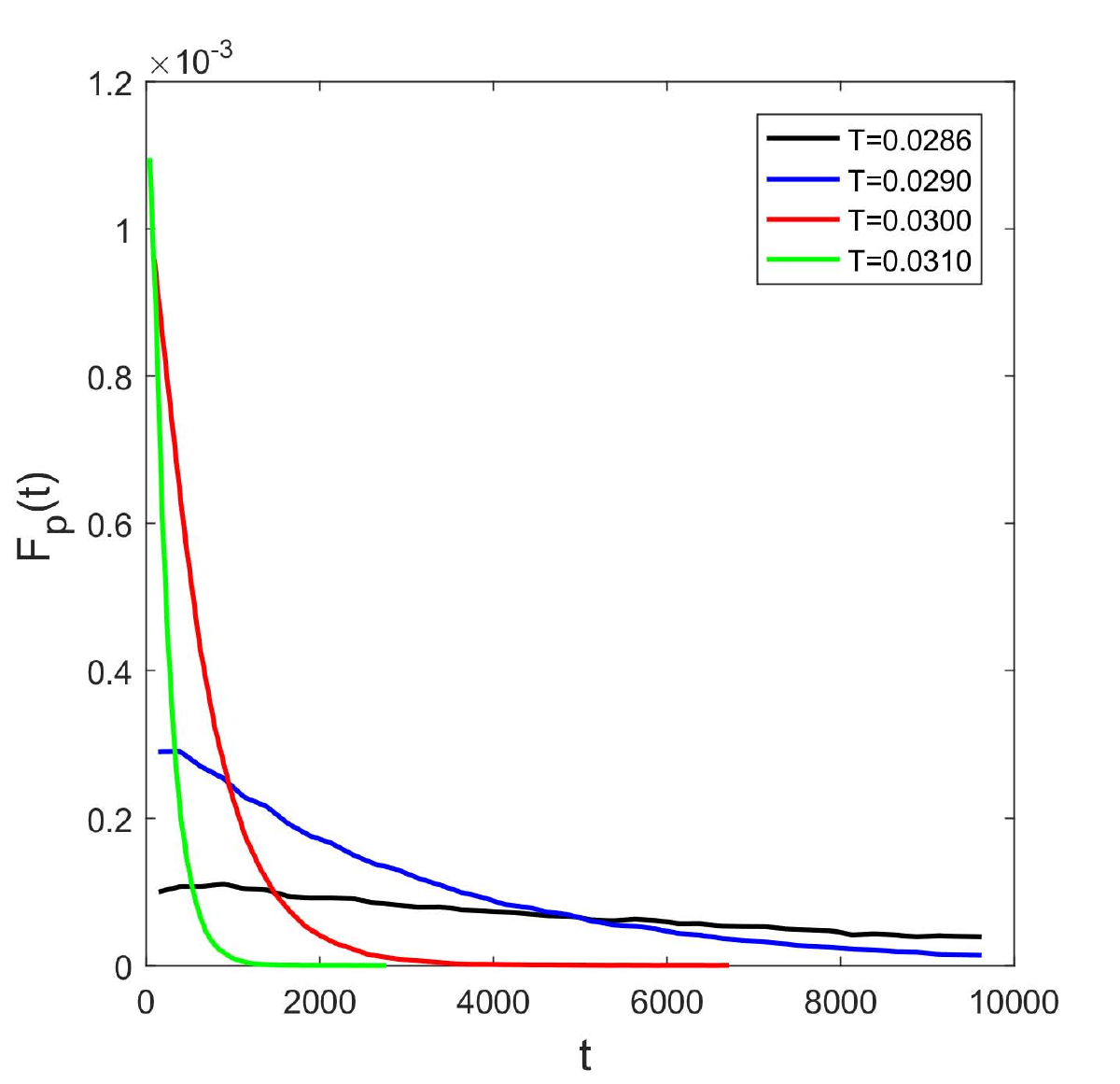}
  \caption{The tail parts of time distributions $F_p(t)$ from small to large black hole state transition at different temperatures. }
  \label{CompareFPTstl}
\end{figure}

In Fig.\ref{FPTdistributionstl}, we display the distribution of first passage time at different temperatures as a function of temperature. The initial distributions are Gaussian wave pockets located at the small black hole state. These plots for different temperatures show the general behavior of the first passage time distribution. It can observed there is a single peak in the first passage time distribution. This implies a considerable fraction of the first passage events occur at short times before the first passage time distribution has attained its exponential decay form. When the temperature increases, the peak becomes more sharper. In addition, the location of the peak moves to the left. This is because the barrier height from the small black hole state to the large black hole state through the intermediate transition state decreases when the temperature increases as observed from Fig.\ref{Barrierheight}. This makes the black hole state easier to go across the barrier under the thermal fluctuation. The tail part of the distribution indicates the fluctuation in kinetics. The longer the tail of the distribution is, the bigger the fluctuations in kinetics are. In Fig.\ref{CompareFPTstl}, the tail parts of time distributions $F_p(t)$ at different temperatures are plotted in one panel. It can be seen the fluctuation will become smaller at higher temperature. This point will be further confirmed by directly computing the fluctuation of first passage time in the following.

With the time distributions, we can calculated the mean first passage time and its fluctuation.
The mean first passage time is defined by
\begin{eqnarray}\label{MFPTcal}
\langle t \rangle=\int_{0}^{+\infty} t F_p(t) dt\;.
\end{eqnarray}
In principle, we can also calculate the $n$-th moment of time distribution function of first passage time by the relation
\begin{eqnarray}\label{nthmoment}
\langle t^n \rangle=\int_{0}^{+\infty} t^n F_P(t) dt\;.
\end{eqnarray}

\begin{figure}
  \centering
  \includegraphics[width=6cm]{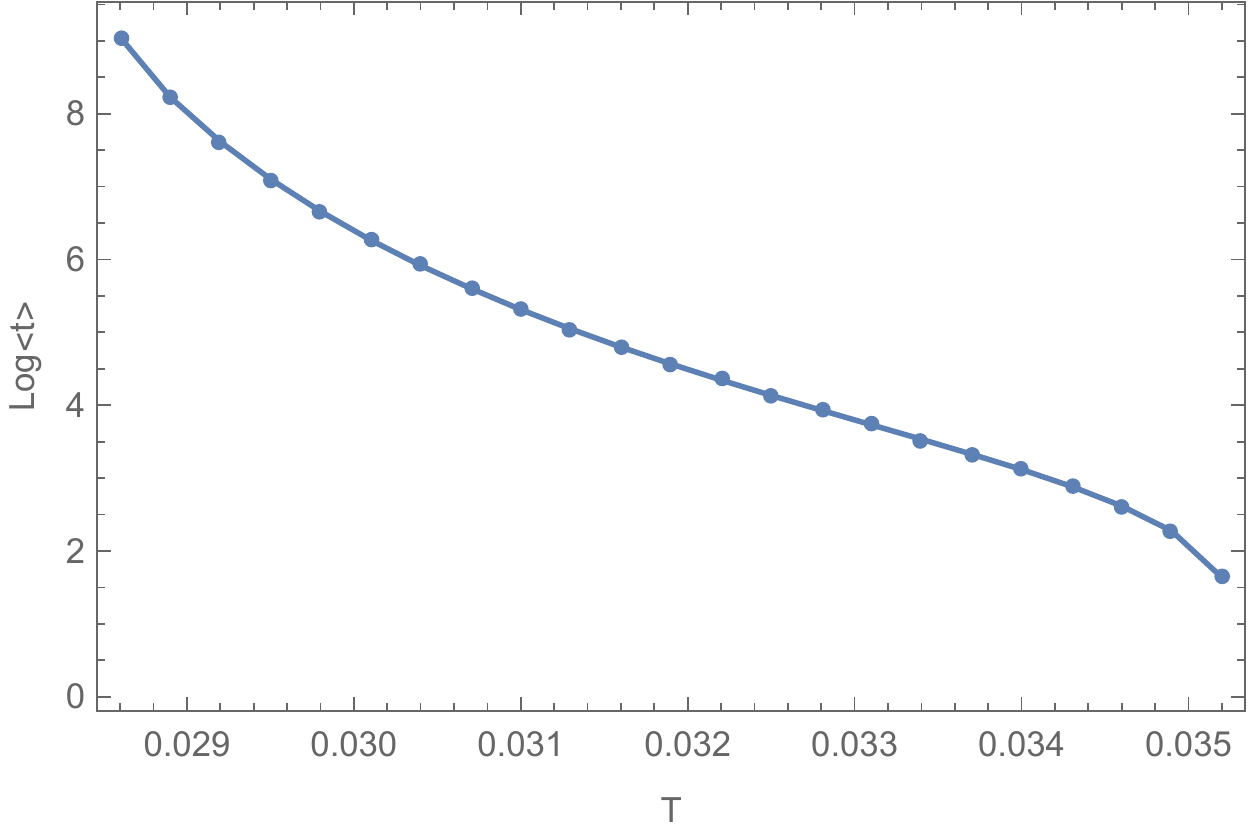}
  \includegraphics[width=6cm]{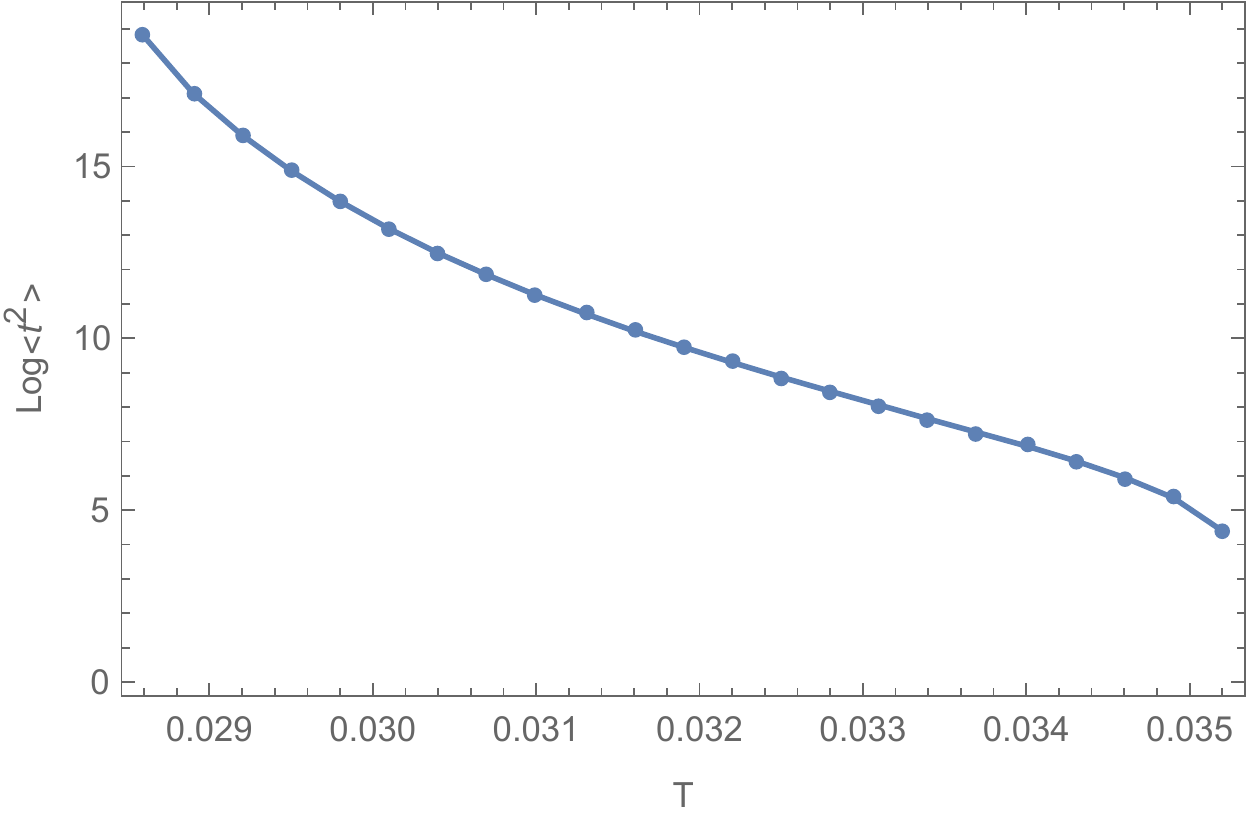}
  \caption{
  The left panel is the plot of mean first passage time $\langle t \rangle$ from small to large black hole state transition as a function of temperature $T$ and the right is the second order moment of the first passage time distribution $\langle t^2 \rangle$ from small to large black hole state transition. The initial distribution is Gaussian wave pocket located at the small black hole state.
  }
  \label{MFPTstl}
\end{figure}

The mean first passage time $\langle t \rangle$ from small to large black hole state transition and the second order moment of time distribution $\langle t^2 \rangle$ are plotted as a function of temperature $T$ in Fig.\ref{MFPTstl}. Note that the vertical axis is in the logarithmic scale. The mean first passage time from the small black hole state to the large black hole state is a monotonic decreasing function of temperature. There are two reasons for this behavior. The first one is that the barrier height from the small black hole state to the large black hole state through the intermediate transition state becomes smaller. Although the thermodynamic stability is determined by the free energy of the black hole state, the kinetics is determined by the barrier height between from the small black hole state to the large black hole state through the intermediate transition state. The second reason is due to the temperature since the thermal diffusion process becomes more effective at higher temperatures.

\begin{figure}
  \centering
  \includegraphics[width=6cm]{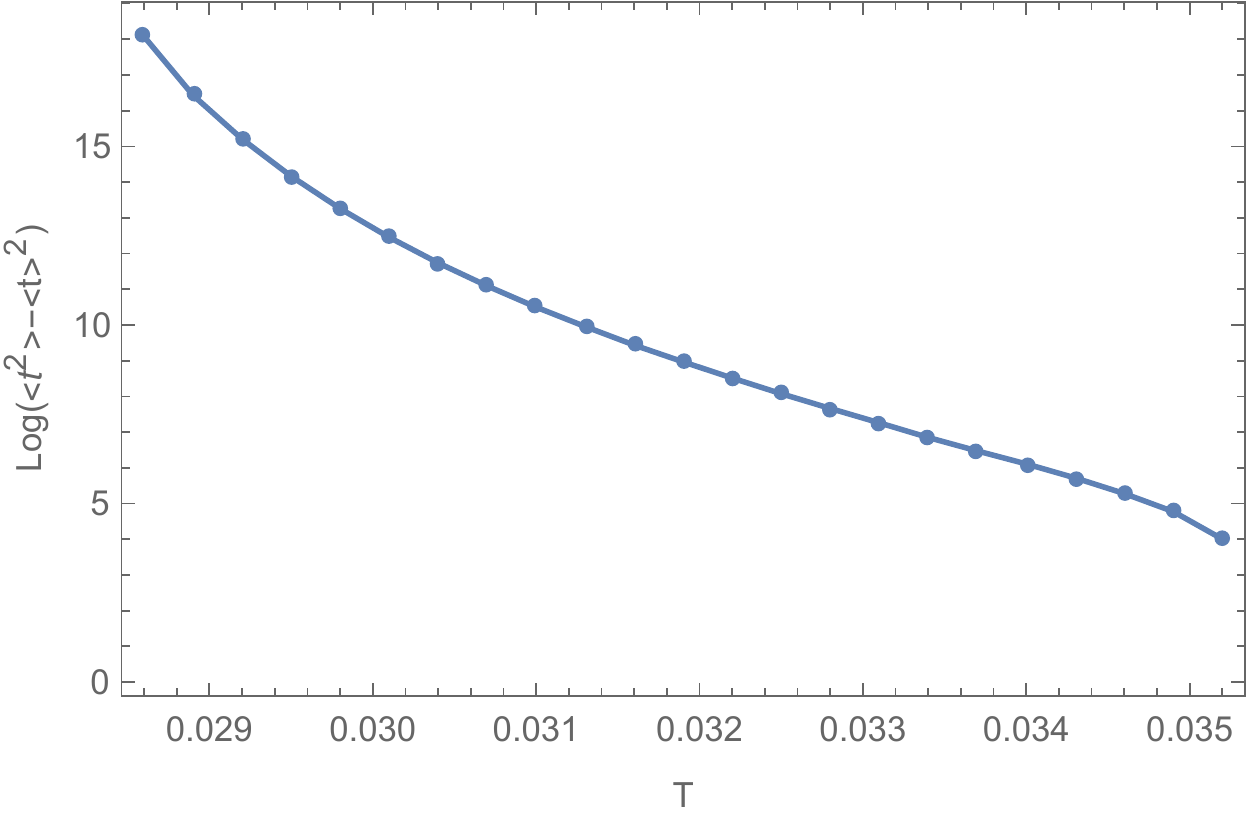}
  \includegraphics[width=6cm]{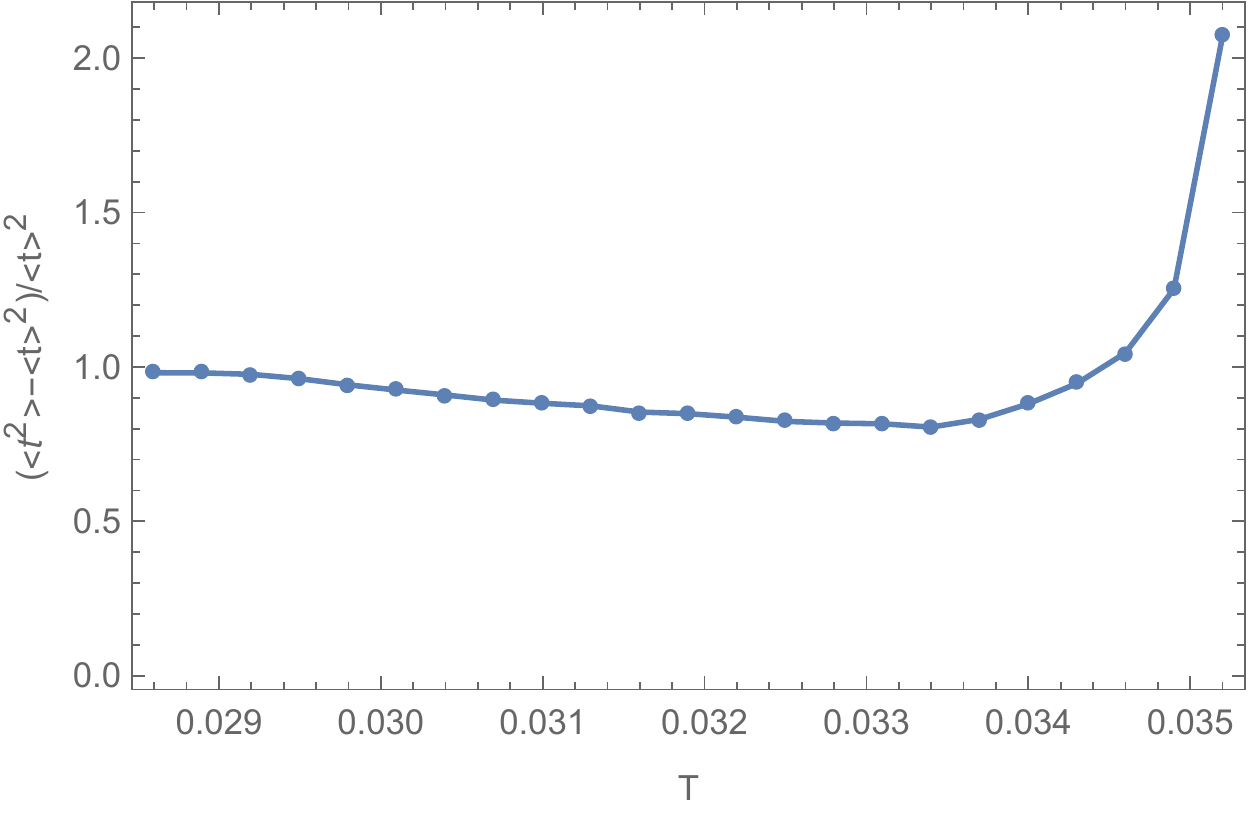}
  \caption{The left panel is the plot of fluctuation $\langle t^2 \rangle-\langle t \rangle^2$ as a function of temperature $T$ and the right is the relative fluctuation $(\langle t^2 \rangle-\langle t \rangle^2)/\langle t \rangle^2$. The initial distribution is Gaussian wave pocket located at the small black hole state.}
  \label{Flucstl}
\end{figure}

The fluctuations and relative fluctuations of first passage time are depicted in Fig.\ref{Flucstl}.
It can be seen fluctuations will decrease when the temperature increases. This behavior is consistent with the distributions of first passage time displayed in Fig.\ref{FPTdistributionstl}, where the peak becomes sharper at higher temperatures. The relative fluctuations decrease firstly when the temperature increases, and attain a minimum at the temperature $T\thickapprox 0.033$. Then the relative fluctuations begin to increase with the temperature. It is obvious that the relative fluctuations attain the maximum at the temperature where the mean first time is at its minimum. We believe that the behavior of the relative fluctuations of the first passage time is the consequence of the two elements mentioned above. As the temperature increases, the free energy barrier height becomes smaller while the temperature becomes higher. The higher thermal fluctuations relative to the barrier height indicate that the thermal fluctuations will have more significant impacts on the kinetics and associated fluctuations than the free energy barrier height at high temperatures. Therefore, the larger thermal fluctuations relative to the free energy barrier height lead to larger relative fluctuations in kinetics at high temperatures. At last, it can be concluded that the effect of temperature dominates.

At last, we will discuss the first passage kinetic process from large black hole to small black hole. This time the time distribution of first passage process is given by
\begin{eqnarray}
F_p(t)=\left.D\frac{\partial}{\partial r}\rho(r,t)\right|_{r=r_m}\;,
\end{eqnarray}
where we impose the reflecting boundary condition at $r=+\infty$ and the absorbing boundary condition at $r=r_m$ (transition state at the free energy barrier top). By numerically solving Fokker-Planck equation with the initial condition and boundary conditions, we can also get the time distributions of first passage process from the large black hole to the small black hole.

\begin{figure}
  \centering
  \includegraphics[width=6cm]{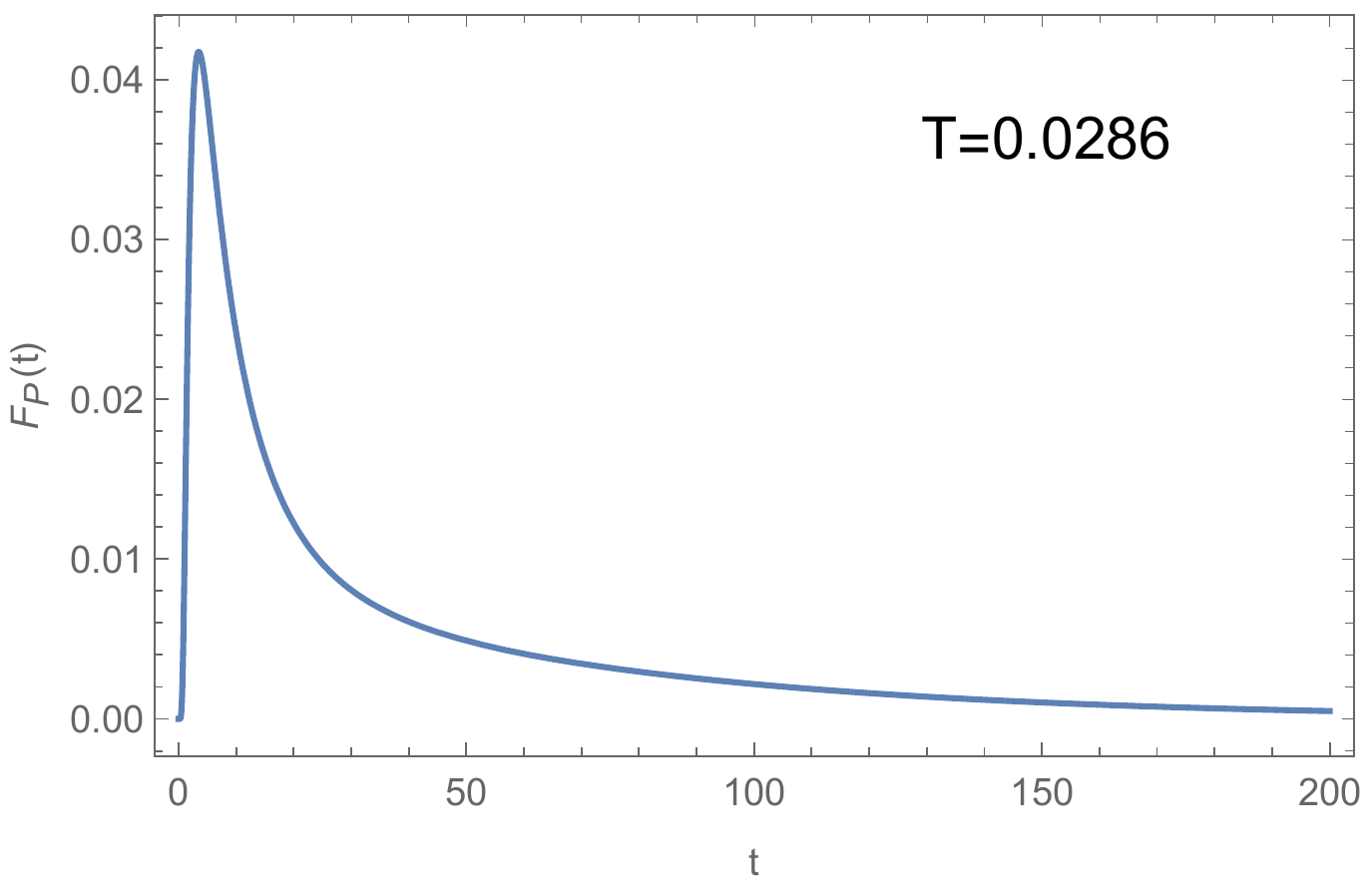}
  \includegraphics[width=6cm]{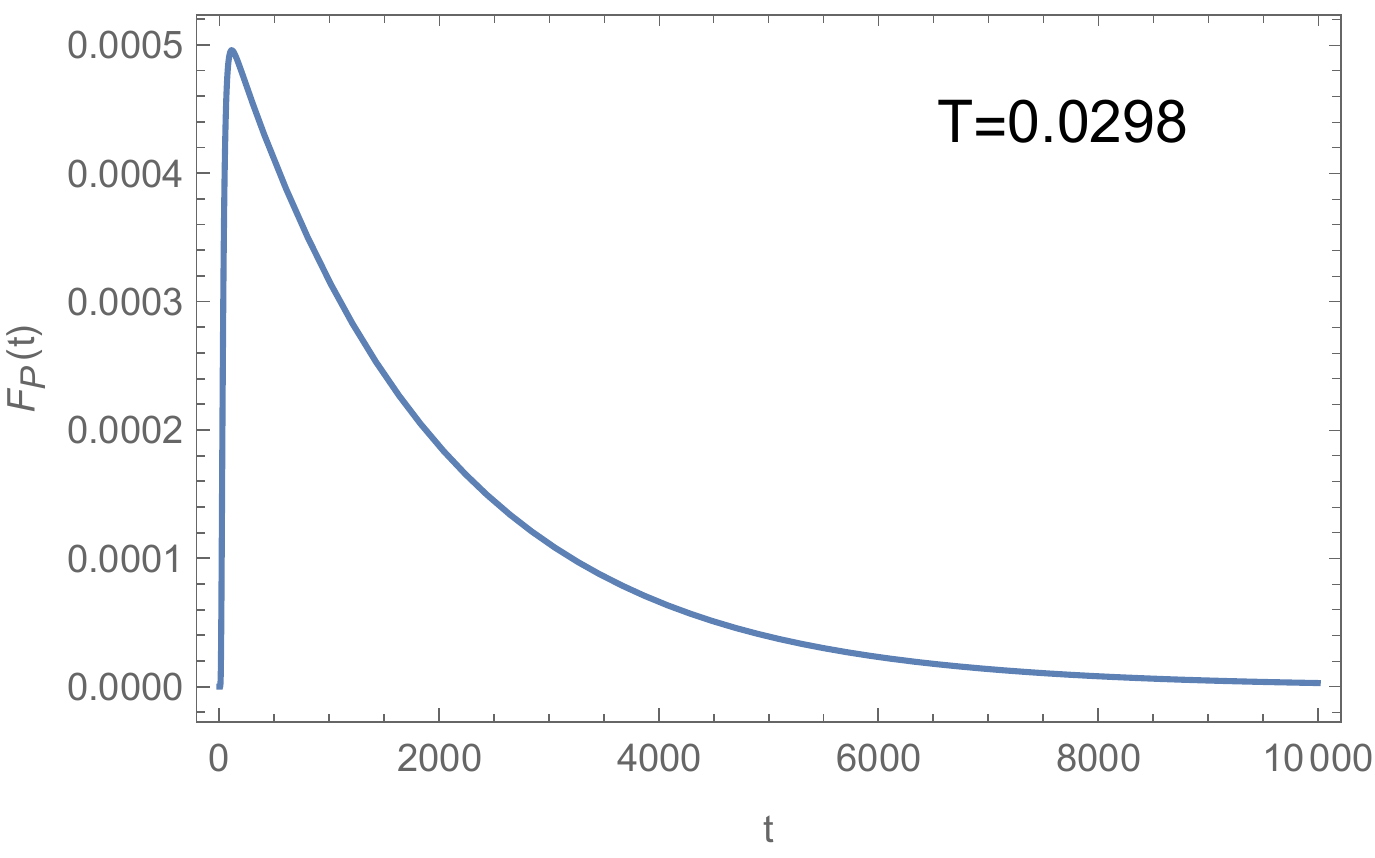}\\
  \includegraphics[width=6cm]{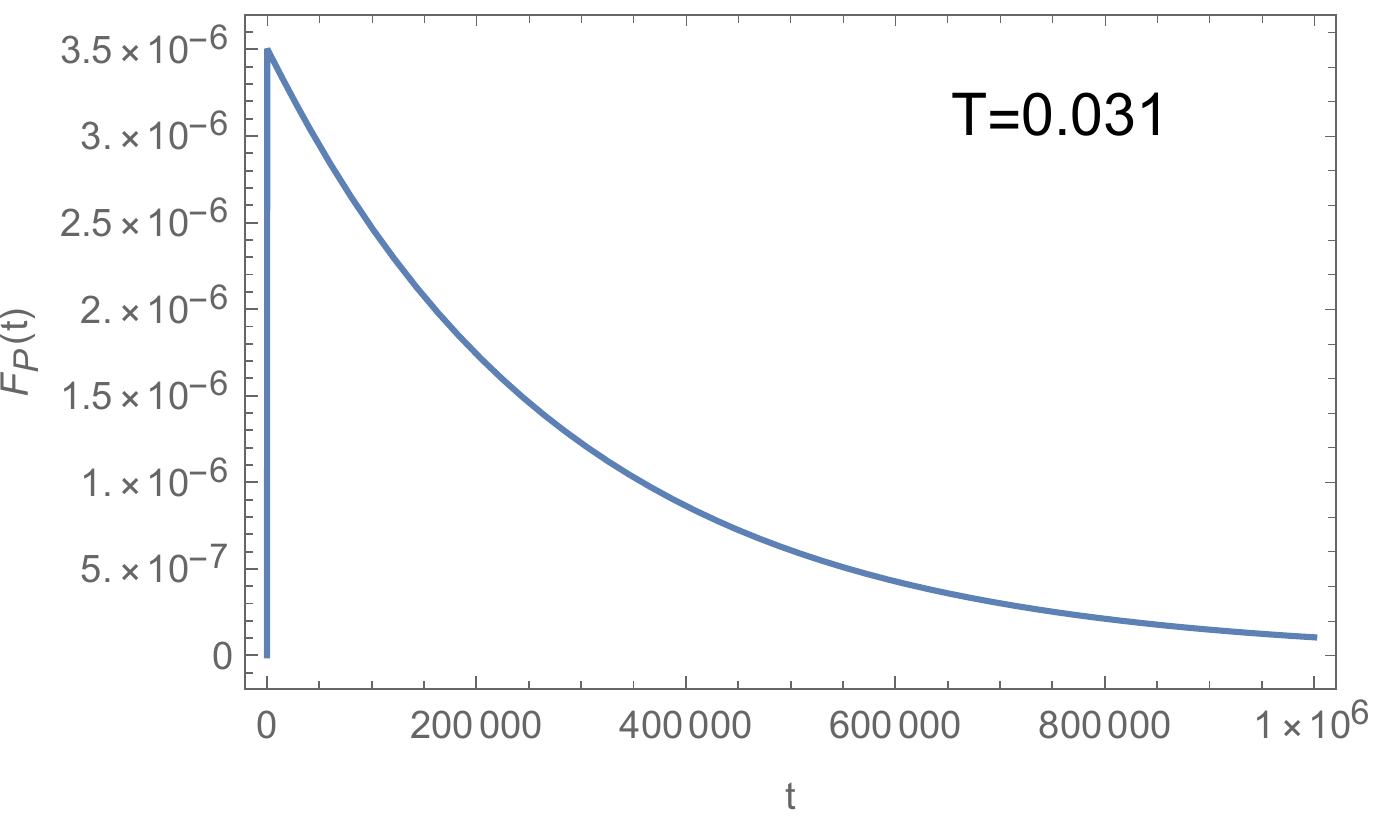}
  \includegraphics[width=6cm]{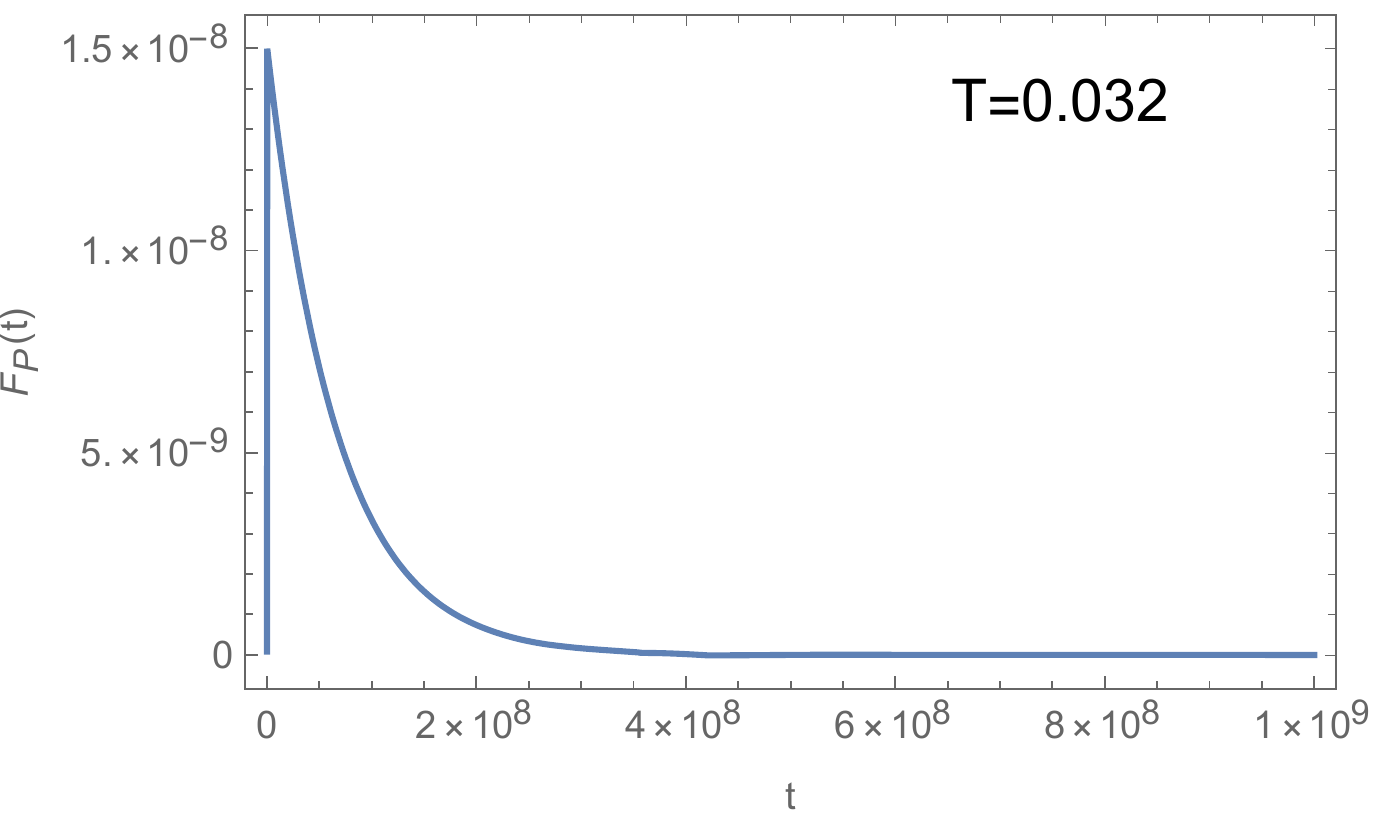}
  \caption{The distributions of first passage time $F_p(t)$ at different temperatures. The initial distribution is Gaussian wave pocket located at the large black hole state. }
  \label{FPTdistributionlts}
\end{figure}

\begin{figure}
  \centering
  \includegraphics[width=6cm]{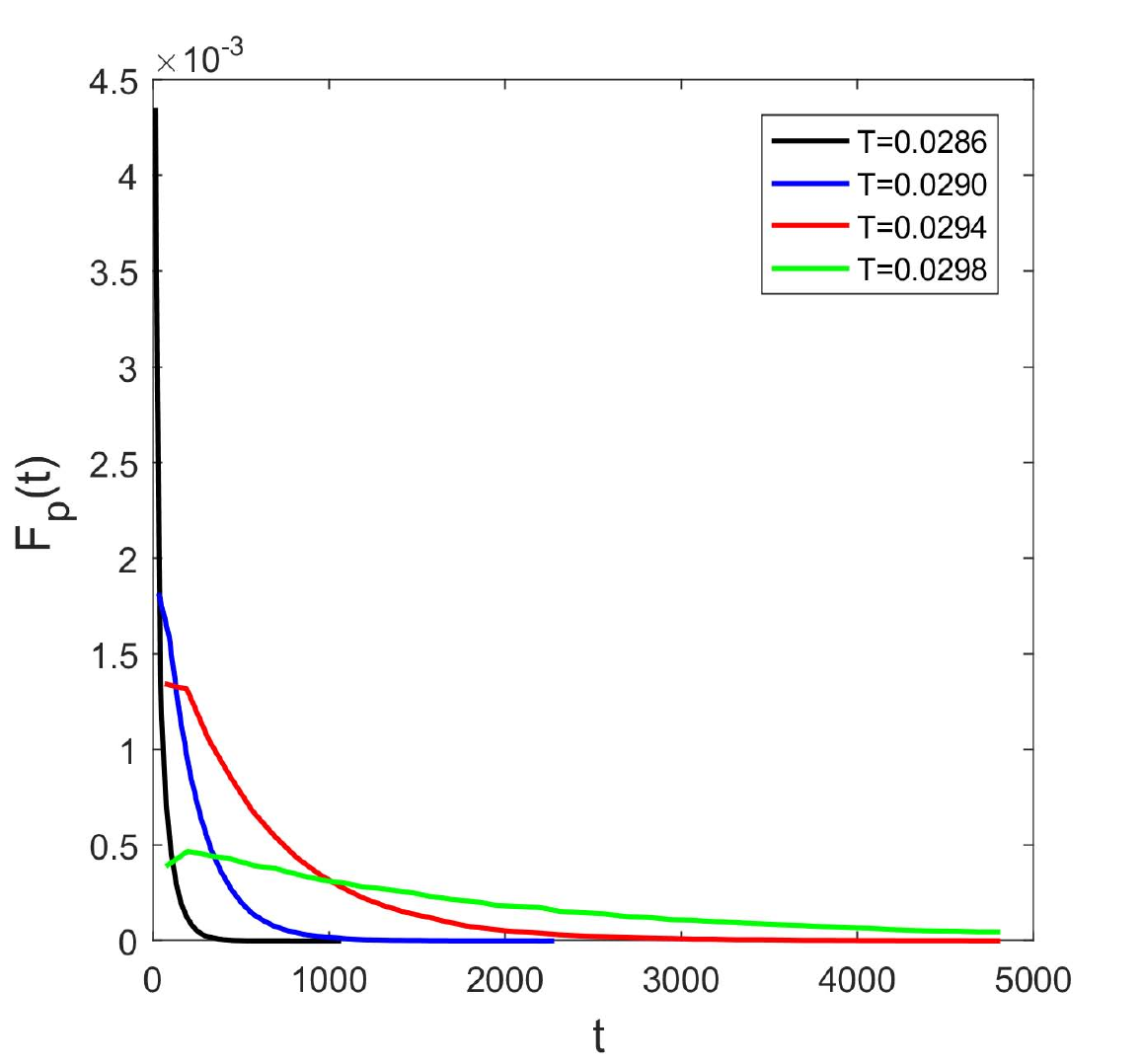}
  \caption{The tail parts of time distributions $F_p(t)$ from large black hole to small black hole at different temperatures. }
  \label{CompareFPTlts}
\end{figure}

In Fig.\ref{FPTdistributionlts}, we display the time distribution of first passage process from large black hole to small black hole at different temperatures. It should be noted that when temperature increases the barrier height from the large black hole to the small black hole on the free energy landscape becomes significantly higher as observed from Fig.\ref{Barrierheight}. It takes the large black hole state extremely long time to complete the first passage kinetic process of switching the small black hole state on average. It is often difficult to simulate the Fokker-Planck equation for such a long time and to control the numerical error. Therefore, we performed the numerical computations from $T=0.0286$ to $T=0.032$. It can observed that there are also a single peak and an exponential decay tail in every time distribution plot. When the temperature increases, the peak becomes wider. We also show the tail parts of time distributions of first passage process from large black hole to small black hole in one panel in Fig.\ref{CompareFPTlts}. It can be seen the fluctuation will become bigger when increasing temperature.

\begin{figure}
  \centering
  \includegraphics[width=6cm]{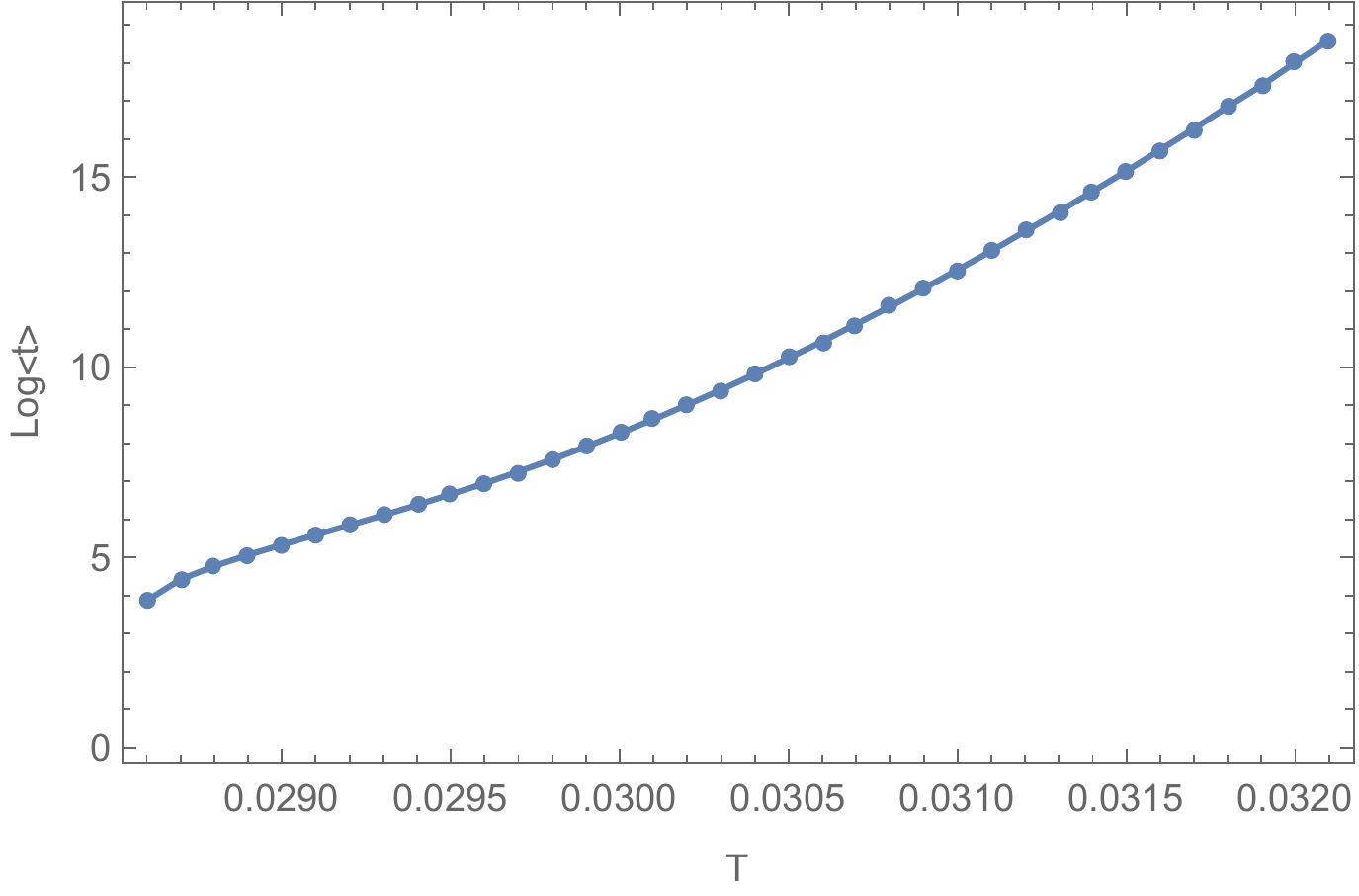}
  \includegraphics[width=6cm]{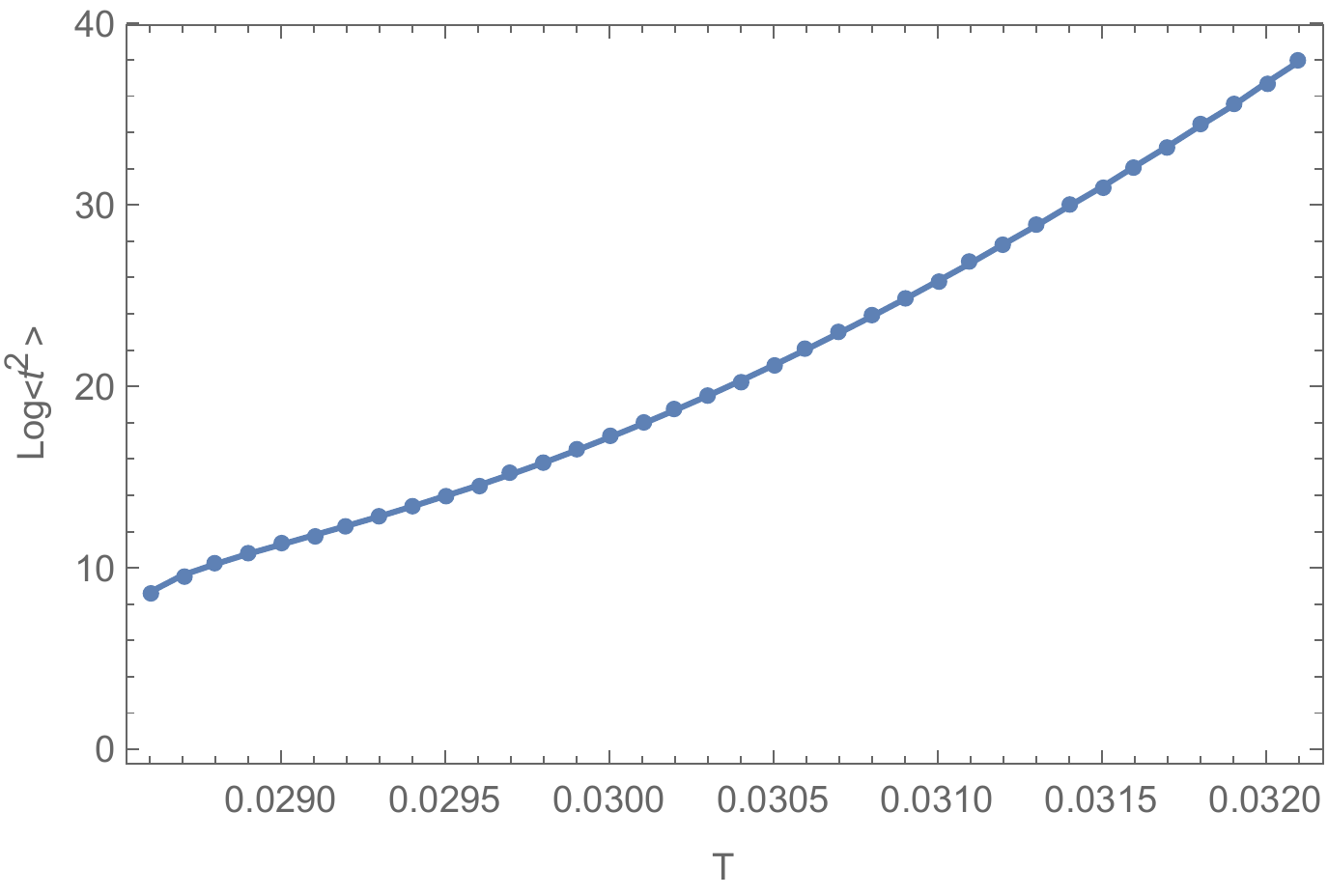}
  \caption{The left panel is the plot of mean first passage time $\langle t \rangle$ from large black hole to small black hole as a function of temperature $T$ and the right is the second order moment of time distribution $\langle t^2 \rangle$. The initial distribution is Gaussian wave pocket located at the large black hole state.}
  \label{MFPTlts}
\end{figure}

\begin{figure}
  \centering
  \includegraphics[width=6cm]{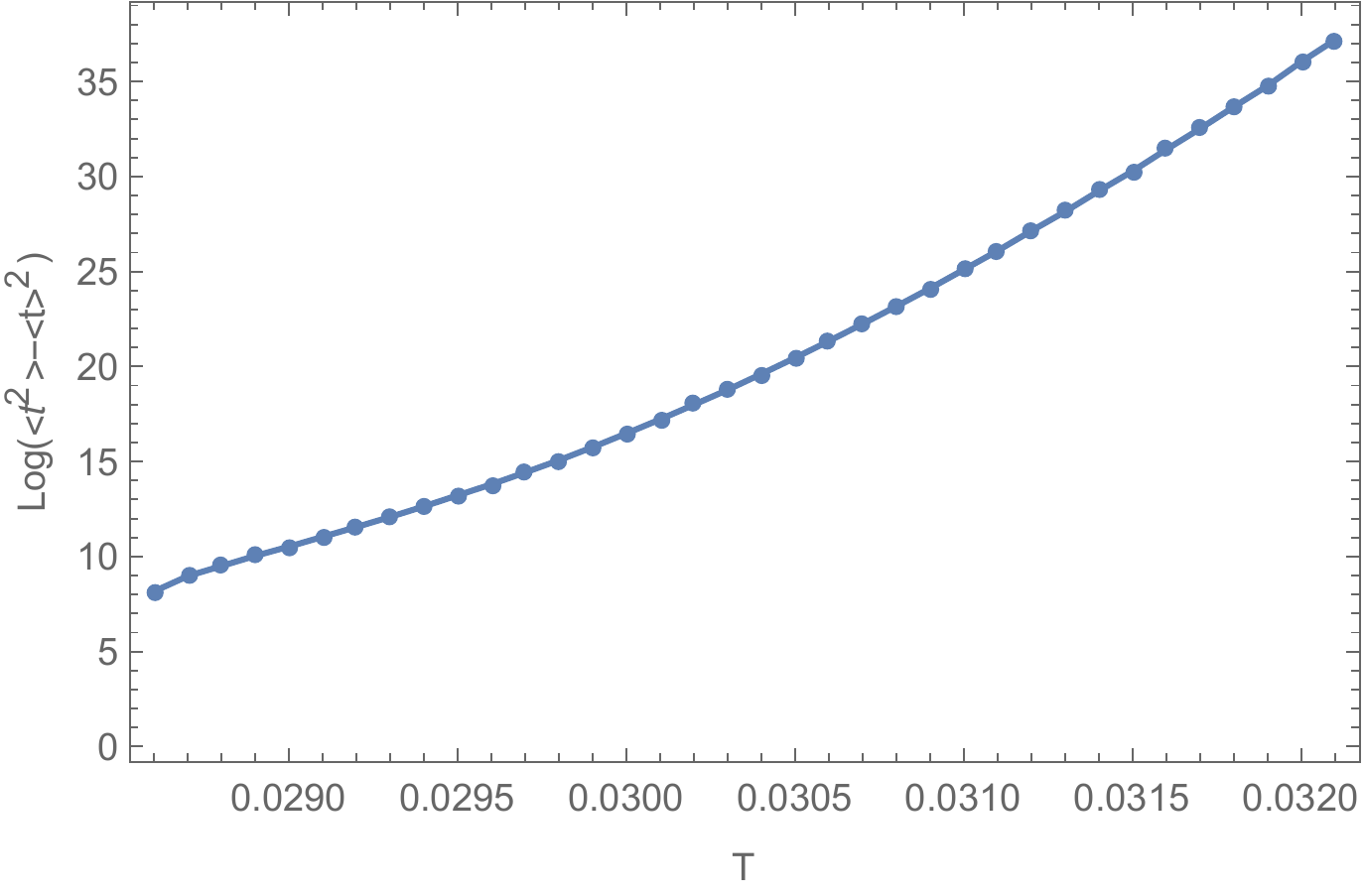}
  \includegraphics[width=6cm]{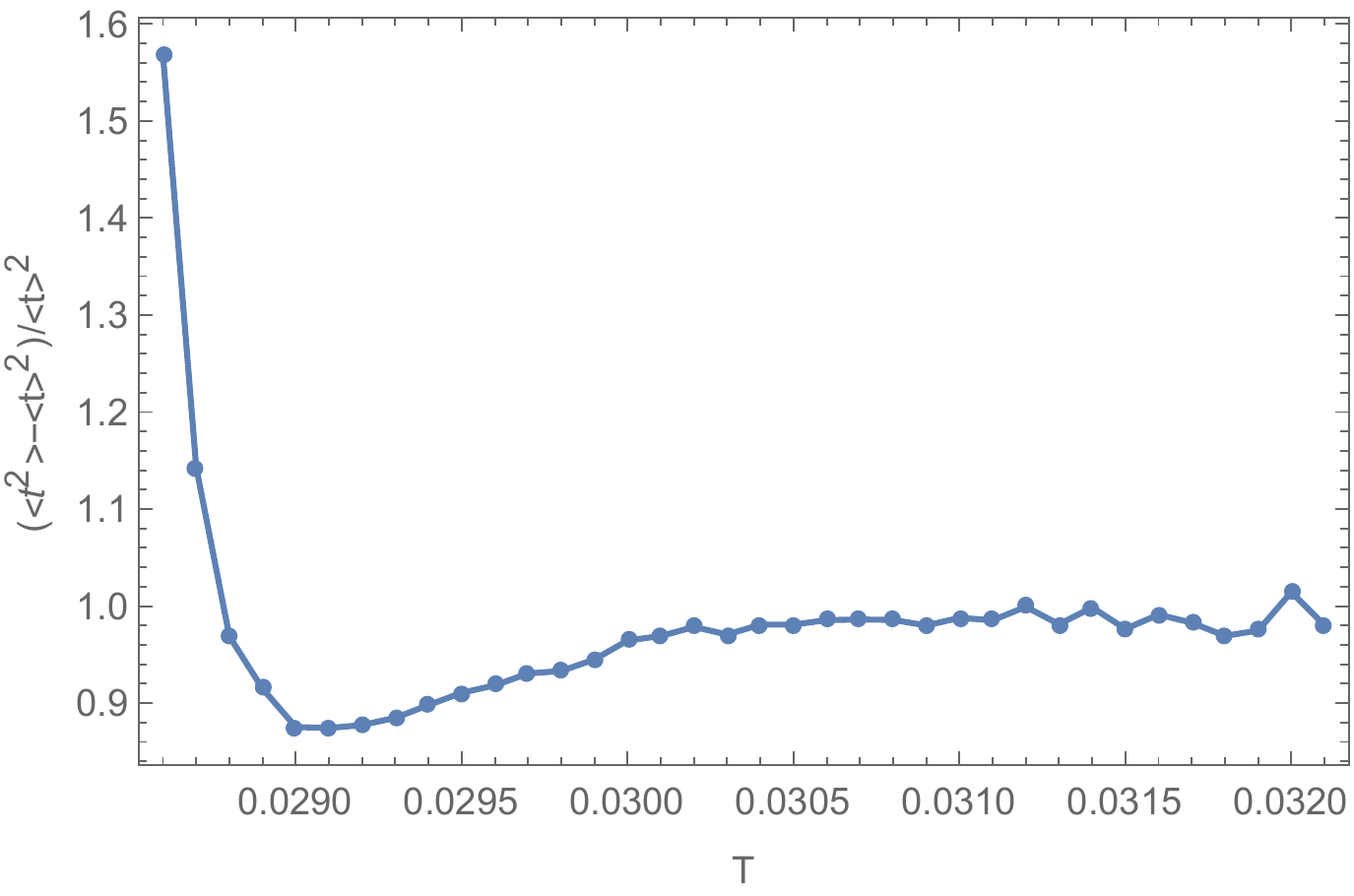}
  \caption{The left panel is the plot of fluctuation $\langle t^2 \rangle-\langle t \rangle^2$ as a function of temperature $T$ and the right is the relative fluctuation $(\langle t^2 \rangle-\langle t \rangle^2)/\langle t \rangle^2$. The initial distribution is Gaussian wave pocket located at the small black hole state.}
  \label{Fluclts}
\end{figure}

The mean $\langle t \rangle$ and the second order moment of time distribution $\langle t^2 \rangle$ of the first passage kinetic process switching from the large black hole to the small black hole are plotted as a function of temperature $T$ in Fig.\ref{MFPTlts}. In this case, the mean first passage time is a monotonic increasing function of temperature. For high temperature, the mean fist passage time becomes very long. This implies that it is very difficult for the large black hole to escape to the small black hole state. We can conclude that the free energy barrier is the dominate factor that impacts the mean time of the first passage kinetic process of switching from the large black hole to the small black hole.

The fluctuations of the first passage time from the large black hole to the small black hole are depicted in the left panel of Fig.\ref{Fluclts}. It can be seen fluctuations increase with temperature. This behavior is also consistent with the distributions of first passage time displayed in Fig.\ref{FPTdistributionlts}, where the peak becomes wider at high temperatures. The behavior of relative fluctuation is plotted in the right panel of Fig.\ref{Fluclts}. The relative fluctuations attain the maximum at the temperature where the mean first time is at its minimum. The conclusion is unchanged in this case. As the temperature decreases, the free energy barrier height relative to the temperature decreases. The higher thermal fluctuations relative to the barrier height illustrate that the thermal fluctuations will have more significant impacts on the kinetics and associated fluctuations than the free energy barrier height at low temperatures. Therefore, the larger thermal fluctuations relative to the free energy barrier height lead to larger relative fluctuations in kinetics at low temperatures. As the temperature increases, the relative fluctuations rapidly decrease from the maximum to the minimum, and then increase slowly. It should be pointed out that, because of the considerable numerical errors at high temperature, it is difficult to compute the relative fluctuation precisely.

\section{Conclusion and discussion}

In summary, we have studied the van der waals type phase transition in RNAdS black holes from the viewpoint of free energy landscape. Black holes are considered as thermodynamic entity. We take RNAdS black holes as a macroscopic emergent state in the extended phase space. Phase diagram show the emergent phases of small and large black holes as well as the coexistence of the small and large black hole as well as the corresponding phase transitions and the associated thermodynamic stabilities. Under thermal dynamic fluctuations, there is a chance for switching from one black hole state to another. The mean first passage time for kinetic switching can be related to the life time of the black hole state. In this spirit, we mainly study the Fokker-Planck equation for the state probability evolution in time on the Gibbs free energy landscape of RNAdS black hole. Then we numerically solve the Fokker-Planck equation and obtain the probability distribution of states and time distribution of first passage kinetic process of black hole state switching.

We emphasize that the concept of the order parameter is very essential to formulate the free energy landscape. As shown, taking the radius of the black hole as the order parameter is intuitive and also convenient to quantify the free energy landscape. This can also help to describe the underlying thermodynamic and kinetic processes at the emergent level. In \cite{WeiLiuPRL,WeiLiuMann}, the concept of black hole molecules is proposed, and the interactions of RNAdS black hole microstructures was studied. Especially, the number density of black hole molecules, which is inversely related to the black hole radius, was proposed to describe the microscopic degree of freedoms. One can expect that if the number density is used as the order parameter, the results and the conclusions will not be changed.

Another assumption we have made is that there exists a series of black hole spacetimes distributed in a wide range of radius at the specific temperature. In the previous studies on the black hole phase transition in the extended phase space, there are three branches of black holes, i.e. the small, the large, and the intermediate black holes. These black holes are distinguished by their radii. Our discussion is based on the free energy landscape, where the free energy is defined as the continuous function of the order parameter. Especially, we are considering both the thermodynamics and the underlying kinetics of the phase transition. For these reasons, besides the small, the large, and the intermediate black hole phases, there should be transient states during the phase transition processes. All these spacetime states compose the canonical ensemble we are considering. The small and the large black hole states are locally or globally stable. Other spacetime states are the unstable transient or excited states, which are bridges for the state switching or phase transition processes. Note that the three branches of the black holes are the on-shell (stationary) solutions of the Einstein field equations while other spacetimes are off-shell and do not satisfy the stationary Einstein field equation. For a specific spacetime we have assumed, the radius of the outer horizon (event horizon) is relevant to our discussion. The inner horizon is always inside of the outer horizon no matter how small the black hole radius is. For this reason, we do not concern about the inner horizon.

The whole physical picture of the black hole phase transition based on the free energy landscape can be understood intuitively as follows. Consider a dynamical equation of $d\vec{x}/dt=\vec{F}(\vec{x})$ where both $\vec{x}$ and $\vec{F}$ are vectors. Here $\vec{x}={x_1, x_2, x_3, \cdots, x_n}$ represents a state characterized by the $n$ dimensional coordinate $(x_1,..., x_n)$. We can think of this equation as the Newtonian dynamics for overdamped system with large frictions so that the usual mass or inertial term can be ignored. Then the $\vec{F}(\vec{x})$ has the physical interpretation of the force. The above equation of motion can be viewed as a particle moving in $N$ dimensional space. Then, the dynamics can be understood as that the evolution of the state is determined by the force $\vec{F}(\vec{x})$. At stationary state or fixed points, $d\vec{x}/dt=0$ and $\vec{F}(\vec{x})=0$. Thus $\vec{F}(\vec{x})=0$ provides the stationary solutions or states of the equation of motion ($d\vec{x}/dt=\vec{F}(\vec{x})$). Suppose that there are several stationary solutions or states corresponding to $\vec{x}_{(1)}$, $\vec{x}_{(2)}$, $\cdots$, $\vec{x}_{(m)}$. Then the dynamics of the state is determined by the force leading to these stationary states depending on their stabilities. Notice that the force in general does not satisfy the equation $\vec{F}(\vec{x})=0$ during the dynamical process.

The analogy here is clear. The Einstein equations, which give the on-shell (the small and the large) black hole solutions, determine the stationary states or fixed points in the dynamical system analogy. However, the evolution of these emergent states and the corresponding state switching are not dictated by the stationary Einstein equations. In other words, the dynamics away from the stationary states has to be off-shell. Since the black holes have temperatures, the thermodynamic fluctuations are inevitable. We can think of the small and the large black holes as the locally stable states while the off-shell black hole states as thermally excited unstable states. Therefore these excited black hole spacetimes are off-shell and considered as the transient states during the dynamical processes. The off-shell dynamics thus represents how the initial spacetime state evolves to the stationary state or how the initial stationary spacetime state can be switched to the other stationary state through the transient states under the thermodynamic fluctuations.

As the black holes have temperatures, their dynamics should be driven by the thermodynamic driving force. In other words, this driving force $\vec{F}(\vec{x})$ for the dynamics should be dictated by the gradient of the Gibbs free energy under the canonical ensemble of black hole states. The Gibbs free energy provides the landscape or weight of the black hole spacetime states either on-shell or off-shell. On the other hand, there are stochastic forces coming from the thermal fluctuations. With both the thermodynamic and the stochastic forces together, the evolution dynamics at the emergent level represented by the black hole radius becomes stochastic and can be described by a stochastic Langevin equation. Therefore, the resulting dynamical trajectories are not predictable. However, the corresponding evolutionary probability of the dynamics satisfies a linear Fokker-Planck equation and is predictable. It is in this sense we can characterize both the thermodynamics through the free energy landscape and kinetics through the evolutionary probability.

It is important to point out that although in principle one can investigate the dynamics of the spacetimes including all underlying degrees of freedom represented, this is in practice not possible. Furthermore, what we really care about is the black hole at the emergent level, their phases and associated phase transitions. We can think about the analogy with the liquid gas phase transition. The underlying degrees of freedom are dictated by the Avogadro's number of molecules. One can in principle simulate the dynamics of all molecules and find out the liquid phase, gas phase and their corresponding transitions. This is in practice not possible. In reality, it is often convenient to use density for representing the degree of freedom at the macroscopic emergent level and distinguish the gas and liquid phase. Therefore, at the emergent level, the density can be used as the order parameter to describe the thermodynamic and kinetic process of liquid phase, gas phase and their corresponding phase transitions. For the same analogy, we will not go after all the underlying spacetime degrees of freedom just as that people usually do not trace all the  molecules for studying the liquid-gas phase transitions. Instead, we will adopt the black hole radius as the order parameter to represent the emergent level degree of freedom for describing the thermodynamics and kinetics of the black holes and black hole phase transitions.

There are still unsolved issues. In the present work, the effect of Hawking radiation is not taken into account. Therefore, it is natural to consider the effect of Hawking radiation on the stochastic thermal dynamic phase transition of RNAdS black holes. The second one is how to relate the first passage time or switching time to the real life time or the stability of the black hole.
This study can help to establish a concrete foundation of studying black hole phase transition from stochastic dynamics viewpoint. At last, the interpretation of the van der Waals type phase transition in RNAdS black holes from the holographic duality viewpoint is still missing in the literature. These questions deserve future studies.

\section*{Acknowledgement}

R. L. would like to thank Dr. Wufu Shi for useful discussion.

 \end{document}